# Evaluation of the Genome Mixture Contents by Means of the Compositional Spectra Method

**Valery Kirzhner**[*1] **& Zeev Volkovich**[2]

[1] Institute of Evolution, University of Haifa, Haifa 31905, Israel;
[2] Software Engineering Department, ORT Braude College of Engineering, Karmiel 21982, Israel.

## Abstract

In this research, we consider a mixture of genome fragments of a certain bacteria set. The problem of mixture separation is studied under the assumption that all the genomes present in the mixture are completely sequenced or are close to those already sequenced. Such assumption is relevant, e.g., in regular observations of ecological or biomedical objects, where the possible set of microorganisms is known and it is only necessary to follow their concentrations.

## 1. Introduction

At present, the most accurate and wide-spread way of detecting a pathogenic bacterium in a living organism is the PCR method, which is based on identifying in the sample a unique DNA (RNA) fragment characteristic only of the targeted bacterium. It should be noted that the technique of PCR requires multiple duplicating of the original DNA molecule. Such process allows detecting even a single bacterium cell in the sample, while estimating the bacterium concentration in the sample appears to be highly problematic.

The metagenomic approach, which has been developed lately, makes it possible to conduct the clinical studies of human microorganisms at an essentially different level. A metagenome is the whole set of the genomes in the bacterial communities living in the functional systems of an organism, e.g., in intestines. Symbiotic relations of the host organism with such communities play an important role in the function and diseases of the human organism. Thus, knowing the contents of such communities is essential in clinical medicine. Since a community may contain a large number of non-cultivated bacteria, it has turned out that the most effective method of investigating a bacterial community is based on analyzing its metagenome *in silico*, in particular, on separating the metagenome into its component genomes with account for their multiplicity (concentration). Such approach is effective, with respect of time costs and expenses, due to dramatical advance in the sequencing techniques, which can now give fast and consistent representation of a metagenome as DNA short fragments. However, at the next step, evaluating different genome fractions in the metagenome takes tens or even hundreds hours of computer time.

---

[*] Corresponding author, e-mail valery@research.haifa.ac.il

In contrast to a regular genome, a metagenome is dynamically changing because the corresponding bacterial community is always under the influence of different randomly varying factors such as nutrition and medicines. For this reason, methods of fast evaluation of metagenomes are essential for regular observations of biomedical or ecological objects. Recently, a mathematical method intended to solve such kind of problems was proposed [1] (see also [2]). This method appears to be especially effective for finding bacteria multiplicities when the metagenome contents is known[1] and it is only necessary to follow the concentrations of bacteria. In this case, the computational time is as short as several seconds or minutes.

In view of the great body of current intensive research, it is obvious that all the genomes that may be present in the metagenomes important for clinical practice will be sequenced in the nearest future. Thus the situation when the metagenome contents is known will become quite common. In this connection, in the present study, we investigate in detail the above-mentioned method [1, 2] as applied to this situation.

The paper consists of seven sections. In Section 2, which follows Introduction, we review the existing methods of metagenome analysis and describe in more detail the method that is most adequate to the problem being solved in this work. In Section 3, the basic model and the possible scenarios of its implementation are discussed. In Section 4, the calculation results are presented and discussed. In Section 5 the prospects for the extension of the model are suggested, while Section 6 is the conclusion.

---

[1] It will be seen from the following analysis that this requirement can (and should) be alleviated.

## 2. Background

In the present paper, the term *genome mixture* refers to a set of DNA segments (i.e., words over a 4-letter alphabet), each segment of the set being part of some genome belonging to a set of genomes, *S*. Such mixture is dealt with in metagenome investigations by *in silico* methods.

There exists a large group of methods in which all the segments constituting the mixture are partitioned into cluster in such a way that each cluster contains only the segments belonging to the same genome. Obviously, the result of this procedure is the *separation of the genome mixture*. These methods employ the distances between the segments which are defined on the basis of different features of the segments such as C+G content, dinucleotide frequencies, and synonymous codons [3-5]. The segments may be also characterized by fixed-length words (see, e.g., [6], where the length of 4 was assumed). It should be noted that, the separation problem being formulated in such a way, the methods of its solution do not require, generally speaking, the knowledge of the genome sequences of the bacteria that constitute the mixture.

The methods of another group, which are, in a sense, opposite to the described above methods, check the presence of a particular microorganism genome in the mixture. Obviously, these methods can be applied only if the genome sequence (possibly, not the whole sequence) is known. The methods that are usually employed in this case search for the similarity between the known genome sequences and the fragments constituting the mixture. Some of these methods, based on the BLAST methodology, use marker genes [7], DNA-polymerase genes [8], and the genes encoding protein families [9]. We will refer to such procedures as *testing the genome mixture*.

Recently, an effective method of the genome mixture separation was proposed [1,2]. The method employs the *compositional spectra (CS)* of genome sequences. The CS method was proposed long time ago [10-14] for the comparison of genomes and/or long genome fragments. By definition [14], the *compositional spectrum* is the frequency distribution of oligonucleotides of length *m* (referred to in the literature as words, *m*-grams, or *m*-mers) which occur in the genome sequence. The existing versions of the method differ mainly in the choice of the set of oligonucleotides, called *support* (*dictionary*), which the frequency distribution (CS) is evaluated for. At present, there exists a large body of research on genome comparisons which employ different versions of this method and produce results indicative of its validity (see, e.g., [15, 16]). Strictly speaking, a set of oligonucleotides produces two different CS ($CS^+$ and $CS^-$) of a genomic sequence depending on the chosen sequence direction ($3' \to 5'$ or $5' \to 3'$, respectively). Usually, the CS is calculated only for one of the two possible directions. Since it is impossible to fix a unique direction for all the fragments constituting a metagenome, the spectrum of each fragment is assumed to be the sum of both $CS^+$ and $CS^-$ spectra: $CS = CS^+ + CS^-$. Such spectrum is referred to as a *barcode* (of the fragment) (for definition, see [17]). Similarly, the spectrum of a genome sequence is also defined as the sum $CS = CS^+ + CS^-$.

The sum of the spectra of all metagenome fragments is, by definition, the metagenome spectrum. It is clear that the metagenome CS is an approximation to the summarized CS of all the genomes which (with regard for multiplicity) constitute the genome pool under investigation. In the present work, CS is considered to be a vector, each coordinate being equal to the sum of the occurrences (and not frequencies) of the corresponding word in the sequence viewed in both directions.

The method proposed in [1] consists in the optimal approximation of the metagenome spectrum by a linear combination of the spectra of known (already sequenced) genomes. This approach implies that if a metagenome consists only of known genomes, each coefficient of the linear approximation should be equal to the number of occurrences of the corresponding bacterium in the metagenome. The authors [1] note that, provided a metagenome contains also unknown genomes, the coefficients should be approximately equal to the sum of the occurrences of the corresponding bacterium and other (unknown) bacteria with relatively close spectra.

In the present research, in the framework of the latter method, we investigate, in detail, the case when a metagenome consists of known genomes and/or unknown ones, which are very close to the known genomes. Such condition is relevant, e.g., in regular observations of ecological or biomedical objects, where it is necessary to follow the changes of the microorganism concentrations in the mixture with time.

Despite the seeming simplicity of the situation, it should be examined with respect to the validity of the results obtained, especially in view of potential medical applications. The source of possible errors lies in the fact that, actually, the metagenome CS differs from the sum of the spectra of all genomes constituting the metagenome. Firstly, since the CS are calculated for individual segments, the words located at the segment junctions in the whole genome are lost; secondly, in the process of metagenome sequencing, some segments of a particular genome may appear under-represented as compared to the content of this genome in the pool. In addition to the possible ill-conditionality of the problem, this may result in significant deviations of the calculated values from the actual concentrations. This effect is also discussed in view of its biological implication.

## 3. Description of method

### 3.1. Techniques employed

*Compositional spectra* are calculated based on all possible 6-letter words. Therefore, the CS vector dimension is 4096 and the value of each coordinate is the total number of the corresponding 6-letter word in the genome sequence regarded in both directions ($3' \rightarrow 5'$ or $5' \rightarrow 3'$).

*Calculation methods*. The evaluation of matrix degeneration and conditionality as well as the solution of linear equation systems was performed using the MatLab standard functions.

## 3.2. The basic model

Consider set $S = \{s_1, s_2, ..., s_m\}$ of the spectra of $m$ different genomes as a set of vectors in linear space $\mathbf{R}^N$, where $N$ is the dimension of the space, which, by definition, equals the number of words in the vocabulary. Let $\sigma = x_1 s_1 + x_2 s_2 + ... + x_m s_m$ be an arbitrary linear combination of these vectors with integer nonnegative coefficients $x$. We refer to vector σ as the mixture of the genome spectra $s_1, s_2, ..., s_m$, coefficients $x_*$ being the multiplicities of each genome occurrence in the mixture. Now the problem of mixture separation can be formulated as finding these coefficients for given vectors $s_1, s_2, ..., s_m$ and vector σ. If the columns of matrix $\mathbf{S}$ are the vectors of set $S$, the problem is reduced to solving the linear equation

$$\mathbf{S}\mathbf{x} = \sigma, \qquad (1)$$

where matrix $\mathbf{S}$ is, generally speaking, a rectangular $N \times m$ matrix ($N > m$) and $\mathbf{x}$ is the vector of variables $x_*$ of dimension $m$. If matrix $\mathbf{S}$ is not degenerate, i.e., vectors $s_1, s_2, ..., s_m$ are linearly independent, the linear system has a single solution. Under this condition, there exists a system of vectors $T = \{t_1, t_2, ..., t_m\}$ which is bi-orthogonal to the system of vectors $S$, which, for a standard scalar product, means that the following equalities are true: $(t_i s_j) = 0$ $(i \neq j)$ and $(t_i s_j) = 1$ $(i = j)$. Then, obviously, $(\sigma, t_i) = x_i$ for any $i = 1, 2, ..., m$.

Let $\mathbf{T}$ be a matrix whose rows are the vectors of set $T$. Then the solution of Eq. 1 can be written in the form:

$$\mathbf{x} = T\sigma. \qquad (2)$$

This formulae is the solution of the mixture separation problem for the case of non-degenerate matrix.

If the method suggested here gives the solution with non-negative coefficients, it coincides with that obtained by the optimization method used in [1]. However, our way of solving the system of equations may give negative coefficients as well. Obviously, small negative coefficients appear as a result of the data noise, while relatively large negative coefficients are indicative of the presence of an unknown genome in the mixture. In can be concluded, thus, that the "direct solution" of the system of equations used by us better reveals the peculiarities of the noise effect, which is important for testing the method.

In the model described above, the same genome set is used both for making up the mixture and for building matrix $\mathbf{S}$. In what follows, these may be, formally speaking, two different sets, which we will refer to as the *mixture set* and the *separating set*, respectively.

### 3.3. Possible scenarios and the solution interpretation with regard for the biological nature of the problem

If system (1) is consistent (which is the condition of the model), the problem of its solution arises when matrix $\mathbf{S}$ is degenerate or ill-posed. In the latter case, the errors in the input data will make the solution quite far from reality. Below we consider the above two possibilities taking into account the data origin.

<u>*Degenerate matrix* $\mathbf{S}$</u>. We will show that the degeneration of matrix $\mathbf{S}$ has a clear biological meaning and thus the results can be interpreted appropriately. In line with [1], it can be claimed that if the number of the genomes under consideration, $m$, is less than the space dimension, $N$ ($m < N$), *there are no biologically significant reasons for the CS vector of one genome to be in the linear span of the CS vectors of the set of some other genomes.*[2] A random occurrence of such vector in this linear span also has a zero probability since the volume of the linear span has a zero measure unless it coincides with the entire space.

However, there is an important exception from the rule formulated above. There may exist a biological reason for the collinearity of the two vectors, namely, the vectors may be considered as collinear if both genomes belong to the strains of the same species. This is the case of more than collinearity - the two vectors are, actually, almost equal to each other since such two genomes have, by definition, only minor differences. We believe that it is hardly possible to imagine other reasons for collinearity of the vectors.

---

[2] If $m > N$, then, for purely formal reasons, $m-N$ spectra are linear combinations of all the others.

Thus, if $N > m$, it can be supposed that, *as a rule, the genome spectra constitute a set of linearly-independent vectors; the only reason for the vectors to be linearly dependent is the coincidence of some of them.* In the latter case, the matrix of system (1) is degenerate as a result of the pair-wise collinearity of some of its columns. With regard to this special type of matrix $S$ degeneration, we suggest the following apparent way of solving the problem. Namely, we reduce matrix $S$ to $S'$, arbitrarily leaving one column in each group of pair-wise collinear ones. Then, if system $\mathbf{Sx} = \sigma$ is resolvable, system $\mathbf{S'x} = \sigma$ has a unique solution, which can be represented using the bi-orthogonal vector set $T$ (see Eq. (2)). Namely, if column $S_i$ of matrix $S'$ had no collinear analogs in matrix $S$, the value of $x_i = (\sigma, t_i)$ is, indeed, equal to the multiplicity of vector $S_i$ occurrence in sum σ. In contrast to this, if column $S_i$ of matrix $S'$ had $p$ collinear analogs in matrix $S$, then

$$x_i = (\sigma, t_i) = C_{1i} x_1 + \ldots + C_{pi} x_p, \tag{3}$$

where the values of $x_1, \ldots, x_p$ are the multiplicities of the corresponding collinear vector occurrences in sum σ, while coefficients $C_{ji}$ depend on the proportion of vector $S_i$ and its *j-th* collinear analog lengths and can be calculated *a priori*. Furthermore, $p$ equations of type (3) can be obtained by choosing , in turn, each of the column of matrix $S$ as a unique representative of the corresponding group of pair-wise collinear columns. Clearly, the solution of the obtained in this way system of equations (4)

$$\begin{aligned} x_1 &= C_{11} x_1 + \ldots + C_{p1} x_p \\ &\ldots \\ x_p &= C_{1p} x_1 + \ldots + C_{pp} x_p \end{aligned} \tag{4}$$

allows to evaluate unambiguously the sums of the occurrences of equal-lengths genomes in the metagenome. This result suggests that the method does not allow discriminating among the bacteria with almost identical genomes, e.g., among different strains of a bacterium species and this fact has a clear physical meaning.

*Conditionality of matrix* $\underline{\mathbf{S}}$. Bad conditionality of a matrix results from the "almost linear dependence" of its columns. In this case, the system of equations has a unique solution, but its evaluation is often a difficult task. For the described-above biological reasons, it can be supposed that this "almost linear dependence" is accounted for by the vectors which we will call "almost collinear". Such CS vectors may appear in genome pairs for some biologically significant reasons, e.g., in the case of evolutionary proximity or, alternatively, co-evolution. However, similar to the collinear vectors considered above, almost collinear vectors still require the genomes to be relatively close, which, in turn, suggests that the spectra lengths are approximately equal. The theory, in this case, is almost the same as the theory for the described above degeneration case. Namely, it can be shown that the solution coordinates which correspond to the vectors lacking almost collinear analogs are stable for data fluctuations, while the coordinates corresponding to almost collinear vectors may depend significantly on the data error. Nevertheless, the same as before, the sums of the coordinates over the whole group of such vectors are stable for data fluctuations.

If the matrix conditionality is so high that it affects the solution precision, "almost collinear vectors" may be selected and dealt with in the same way as described above for the collinear vectors. Namely, to build a system of bi-orthogonal vectors, only one vector of each pair (group) can be used. This will cause the decrease of the conditionality and the obtained occurrence coefficient will be the sum of the multiplicities of all the bacteria of this group. Of course, in the solution there will be an error, the less being the angle between the "almost collinear vectors", the less being the error.

In conclusion of this section, it should be noted that, the genomes of the mixture set and of the separating set being given, it is possible to *a priori* obtain the characteristics of matrix $S$, in particular, its rank and conditionality. Calculating the pairwise scalar products of the vectors of a given set $S$, it is possible to obtain information on their collinearity and *a priori* develop an adequate scheme of solution and assessing the result. In particular, it is possible to conduct simulations in order to evaluate the level of the solution error. As an example, Fig. 1 demonstrates the distribution of the cosines values for the angles between all possible CS pairs for approximately 1300 bacterial genomes[3].

---

[3] Bacterial genomes are obtained from the site http://www.ncbi.nlm.nih.gov/genomes/lproks.cgi .

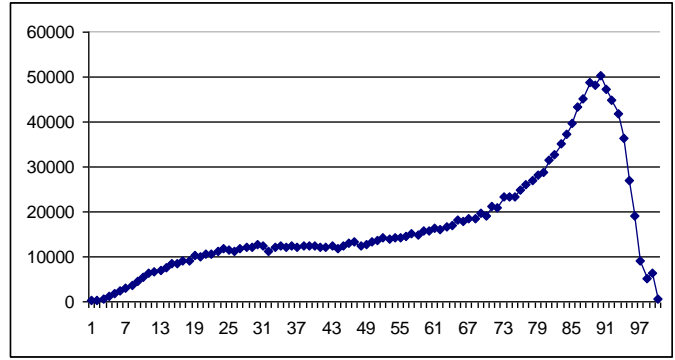

**Fig. 1**. **Distribution of the cosines values for the angles between all possible CS pairs for approximately 1300 bacterial genomes.** X-axis: cosines values ×100; Y-axis: the number of cosine values.

From the data presented in Fig.1, it can be concluded that the number of "almost collinear" vectors is relatively small. The corresponding matrix composed of CS for all considered genomes is not degenerate, so, indeed, the genome compositional spectra do not belong to the subspaces generated by the CS of other genome sets. The conditionality of this matrix equals 545. The contribution of vector pairs with high degree of collinearity to this value can be estimated by calculating the conditionalities of the matrixes in which the vectors forming such pairs are eliminated. For example, eliminating one vector in each pair with the cosine values higher than 0.95, 0.98, or 0.99, we obtain three matrixes with conditionality values of 74, 199, or 228, respectively. Thus the conditionality values appear to be so high that checking the solution accuracy is required; on the other hand, they are quite compatible with the possibility to solve the problem.

## 4. Results and Discussion

### 4.1. Testing the basic model and separation of the mixture in the absence of randomness

The genomic base. To illustrate the calculations in the framework of the described-above model (1), we have considered two sets of genomes. One of the sets, $M_{100}$, contains 100 genomes of Eubacteria, which represent all the main bacteria groups, the number of genomes in each group being approximately proportional to the number of the sequenced genomes in each group. The choice of genomes from each group is random (see

Supplement Informations, Table SI1). The other set, $M_{28}$, consists of 28 bacteria, which are characterized as the most common gut bacteria (see the supplement material in [18]) and, in contrast to other bacteria mentioned in [18], have been completely sequenced by now (Table SI2).

For CS calculations we use all possible 6-letter words, so that the dimension of the full CS space is equal to 4096 (N = 4096). In this way (as it was shown in Section 1) matrices $\mathbf{S}_{100}$ and $\mathbf{S}_{28}$ are created, their dimensions being $4096 \times 100$ and $4096 \times 28$, respectively.

The mixture model. We suppose that each genome that is present in the mixture is cut into non-overlapping segments of equal length and that the mixture is composed of such segments. The spectrum of a genome mixture is defined as the sum of the spectra of all segments. We have considered mixtures composed of segments of length C = 10, 20, 30, 40, 50, 100, 200, 500, 1000, 10000 bp and also, for the sake of comparison, a mixture that consists of whole genomes. The multiplicities of the genome occurrences in the mixture are chosen randomly in the range of 0-10, once for all the numerical experiments in this research.

Direct calculation of multiplicity. Our calculations show that both matrices $\mathbf{S}_{100}$ and $\mathbf{S}_{28}$ are non-degenerate. The conditionalities of matrices $\mathbf{S}_{100}$ and $\mathbf{S}_{28}$ are equal to 314.05 and 78, respectively. However, the relatively high conditionality of matrix $\mathbf{S}_{100}$ does not interfere with the possibility of obtaining an almost exact solution of the corresponding system of linear equations in the absence of noise that is not related to the natural computational errors. For example, if a segment is equal to a whole genome (i.e., the mixture spectrum is calculated accurately), the mean deviation from the actual multiplicity value is 0.00179. Table SI3 presents the results of the calculations of the genome multiplicities in the mixture for different segment lengths and Fig. 2 shows the mean differences between the calculated and the actual genome multiplicities in the mixture.

It has been explained above that linear combinations of spectra do not create new spectra, so the poor conditionality of matrix $S_{100}$ may result from the "almost collinearity" of some spectra. The latter suggestion can be checked by calculating the cosines of the angles between the vectors (Fig. 2). Although most of the coefficients are not close to 1, we have found a few coefficients to be close to 1.

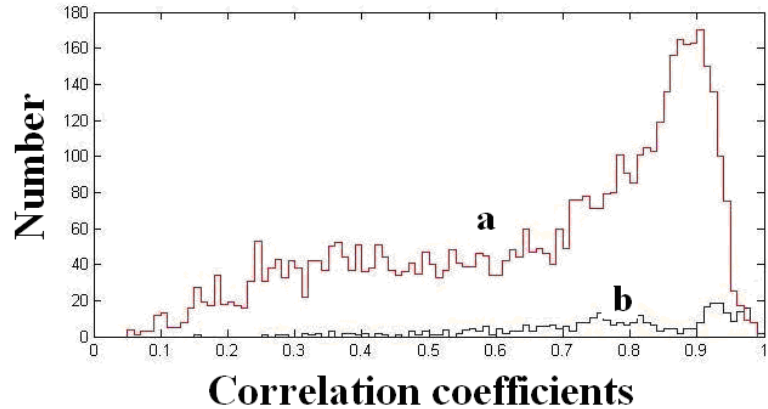

**Fig.2. Distribution of the cosines of the angles between all possible vector pairs.** The number of genomes in the set: **(a)** 100; **(b)** 28.

From the data presented in Table 1, it can be seen that if almost collinear vectors are eliminated, matrix $M_{100}$ becomes much more stable. For example, the elimination of 6 genomes results in approximately a 10-fold decrease of the conditionality.

| 1 | 2 | 3 | 4 | 5 | 6 | 7 |
|---|---|---|---|---|---|---|
| 36, **75** | Mycobacterium bovis | M. tuberculosis F11 | 0.99991 | 4345 | 4424 | 288 |
| **28**, 42 | S.pyogenes | S.pyogenes SSI-1 | 0.998939 | 1841 | 1894 | 285 |
| **95**, 96 | H.influenzae R2846 | H. influenzae R2866 | 0.998936 | 1819 | 1932 | 283 |
| **18**, 25 | L. monocytogenes str. 4b F2365 | L. monocytogenes strain EGD | 0.998768 | 2905 | 2944 | 281 |
| **22**, 48 | S. aureus RF122 | S. aureus strain MSSA476 | 0.998579 | 2742 | 2799 | 158 |
| **12**, 53 | X. axonopodis | X. campestris | 0.995408 | 5175 | 5148 | 157 |

**Table 1. The most collinear bacteria pairs from set $M_{100}$, arranged in descending order with respect to the collinearity value.** 1 – the numbers of bacteria on the entire list (Table A1); 2,3 - the names of bacteria; 4 – the values of the cosines of the angles between the vectors; 5,6 – genome lengths (x1000); 7 – conditionalities of matrix $S_{100}$ calculated after the genomes marked in bold and located not lower than the corresponding row have been eliminated from the entire set $M_{100}$. For example, for the 1st row, the conditionality is calculated for set $M_{100}$ without genome

number 75; for the 2$^{nd}$ row, the conditionality is calculated for set $M_{100}$ without genomes number 75 and 28.

Since the $M_{28}$ genome set conditionality is good enough for performing calculations, it can be supposed that the angle between the vectors in the almost collinear genome pairs is much larger in this case. Indeed, only for one genome pair (*E. coli - E. fergusonii*), the cosine value is 0.993 and there are only two other values slightly exceeding 0.98. With $M_{28}$ set as both the separating and the mixture set, the calculated mean deviation of the obtained multiplicity from the actual one is 0.04097 if the segment length in the mixture is equal to the genome length. The calculated genome multiplicities for different segment lengths are presented in Table SI4, while Fig. 2 shows the mean differences between the calculated and the actual genome multiplicities in the mixture.

<u>Reduction of the separating set</u>. Now let us employ another calculation method, which consists in eliminating one vector from each pair of almost collinear vectors of set $\mathbf{S}_{100}$ (marked in bold in Table 1). The remaining 94 genomes constitute a separating set $\mathbf{S}_{94}$. Employing this set, we cannot calculate separately the multiplicities of the occurrences in the mixture of both genomes (the remaining and the eliminated ones) of the almost collinear pair. The calculated multiplicity of the remaining genome of each almost collinear genome pair is equal to the sum of the multiplicities of the genome itself and the genome lacking from this pair. For example, consider the pair of almost collinear *M. Bovis* and *M. tuberculosis* genomes (Table 1). Elimination of the latter genome from the separating set results in the *M. Bovis* multiplicities equal to 7.2417, 7.9169, and 7.3478 with the segment lengths of 10, 20 and 30, respectively, while the actual summarized multiplicity is equal to 7. The mean difference between the calculated and the actual genome multiplicities in the mixture is shown in Fig. 3.

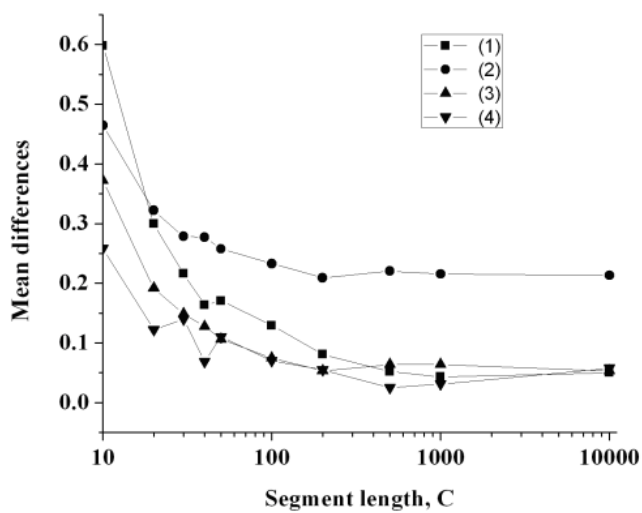

**Fig. 3. The mean differences between the calculated and the actual genome multiplicities in the mixture as a function of the segment length** (log scale is used for the *x*-axis) for sets $M_{100}$ and $M_{28}$. The mixture is composed of: (1) the whole set $M_{100}$ and the separating matrix contains all the genomes; (2) the whole set $M_{100}$, but the separating matrix contains only one genome of each almost collinear pair. The mean differences are obtained based on the difference between the calculated (non-integer) and the actual multiplicity; (3) the same as in (2), but the obtained multiplicity is approximated to the nearest integer; (4) the whole set $M_{28}$ and the separating matrix contains all the genomes.

Noise effect. Next, in order to demonstrate the effect of matrix $\mathbf{S}_{100}$ bad conditionality on the errors in calculating the multiplicities, we have performed the calculations with regard for the noise introduced into the mixture vector. Into each coordinate of the accurate spectra, noise was introduced, which was randomly and evenly distributed between 0% and 1% of the coordinate value. As a result, the calculated multiplicity values for the most collinear genome pair, *M. bovis - M. tuberculosis* (see Table 1), are 7.14 and 0.03 as compared to the actual values of 4 and 3, respectively. However, the sums of the calculated (7.17) and the actual (7.0) multiplicities are much closer to each other, in accordance with the above considerations. The next two pairs of almost collinear genomes in Table 1 are also subject to the introduced error (Table 2).

| 1 | 2 | 3 | 4 |
|---|---|---|---|
| 28 | 2 | 1.9944 | 1.639 |
| 42 | 7 | 7.003 | 7.225 |
| sum | 9 | 8.9944 | 8.864 |
|  |  |  |  |
| 95 | 1 | 1.0005 | 0.443 |
| 96 | 4 | 4.0012 | 4.539 |
| sum | 5 | 5.0017 | 4.982 |

**Table 2. The values of multiplicities calculated in the absence and in the presence of noise as well as the actual values for both pairs**. 1 – genome numbers; 2 – actual multiplicity values and their sums; 3 - calculated multiplicity values in the absence of noise; 4 – calculated multiplicity values in the presence of noise.

The case when separating and mixture sets are different. Consider set $M_{11}$, consisting of 11 different E. coli genomes. The correlation coefficient between each pair of these genomes is larger than 0.99. Let this set be the mixture set and the separating set be set $M_{100}$, which contains only one E. coli genome. The separation obtained for the mixture of the whole genome spectra is presented in Fig. 4.

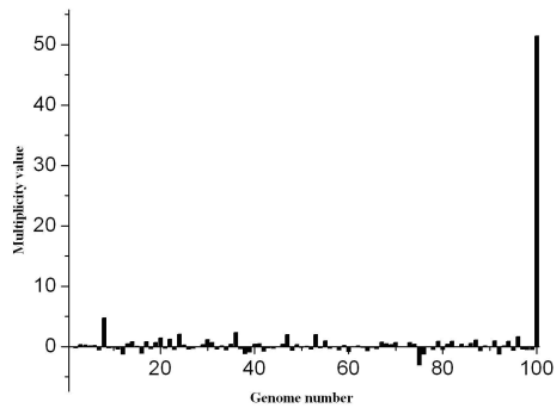

**Fig. 4. Histogram of the expansion coefficients for the set of 11 *E.coli* genomes** over the set of 100 genomes, one of these being also a *E.coli* genome.

The calculated total coefficient for the E. coli genome is 50, while the actual one is 64. The other coefficients are not equal to zero, but almost all of them are less than 1 (see Fig. 4). The largest coefficient, equal to 4, corresponds to *Salmonella* (number 8 in Table A1), which can be readily understood from the biological point of view.

Consideration of more examples of this issue, i.e., the sets that consist of 200, 500, or 1000 genomes, can hardly clarify the situation any further. It can be expected that with the increase of the genome number, the probability of the occurrence of collinear and almost collinear pairs also increases, which, in turn, increases the conditionality of the system. At the same time, since we are considering the properties of already known genomes, all of the above collinearity possibilities can be tested directly.

**4.2. Separation of a mixture with random fluctuations**

In this section, we use the following simple model for random generation of a metagenome spectrum.

Model of metagenome random fluctuation and normalization of the result. Consider again genome sets $M_{100}$ and $M_{28}$. We use the same, as in the previous section, integer coefficients $x$, but the genome spectrum is calculated in a different way. Namely, we include each genome segment into the mixture with an integer value of multiplicity, distributed evenly from 0 to the fixed value $x$ for this genome. The idea of this model is that, actually, not all the segments, but only some random portion of them, are present in the sequenced metagenome. For both sets $M_{100}$ and $M_{28}$, we have conducted the model simulation 100 times for the same segment lengths that were used before.

It should be noted that, in contrast to the deterministic case considered above, in the framework of our probabilistic model, the solution of Eq. 1 fundamentally cannot give even the approximate actual multiplicity of a genome in the mixture. The reason for this is that the described procedure efficiently decreases this multiplicity to the level which is determined by the properties of the randomizing process. Although pair-wise multiplicity ratios are preserved, the calculated absolute values must be lower than the actual ones. Assuming different properties of the process of selecting the mixture segments, it is possible to introduce different recovery coefficients. We, however, propose a simple technique of normalizing the result, which lies a little bit away from pure theory. Namely, prior to metagenome sequencing, a known number of one or two bacteria species must be added to the metagenome. It is desirable that these bacteria be, in biological terms, as far as possible from the supposed composition of the metagenome. Then the ratio of the known multiplicity of each of these bacteria to the calculated multiplicity will be the sought for proportion coefficient for all the bacteria in the mixture. In the following computer experiments, we consider the first genome on the list to be such an added genome. The same trick can be successfully used in the estimation of the inaccuracy caused by the ill-conditionality of the system.

Experiments with the fluctuation model. The characteristics calculated in this case are the mean multiplicity value $d_i$ ($i=1,…,100$) for each bacterium and the squared deviation $\sigma_i$ for each $d_i$ (Tables SI5, SI6) (averaging is performed over 100 experiments in each series). Calculating deviations $d_i$ from the corresponding actual multiplicities and averaging these values over all bacteria, we can assess the quality of solving the mixture-separation problem at different segment length values in the mixture (Fig. 5).

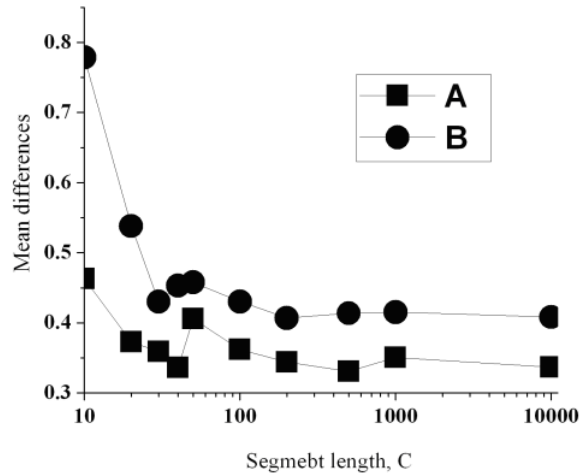

**Fig. 5. The dependence of the mean error in evaluating the genome multiplicities in a mixture on the segment length** (log scale is used for the *x*-axis) for genome sets $M_{100}$ (circles) and $M_{28}$ (squars).

From the data presented in Fig. 5, it can be seen that different segment lengths result in different mean errors, the dependence being non-monotonous. The mean values of the mean-squared deviation are shown in Fig. 6. On the whole, this characteristic increases at the ends of the segment-length ranges.

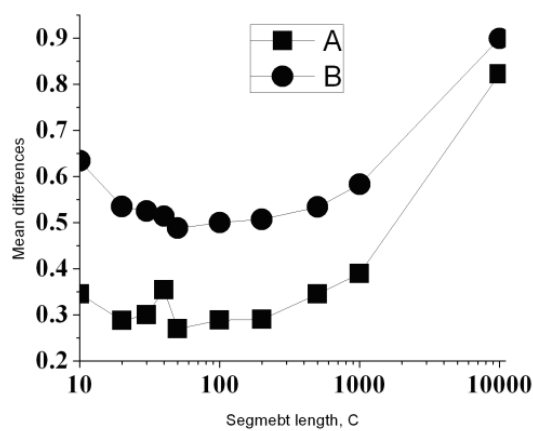

**Fig. 6. The dependence of the mean-squared deviation of the genome multiplicities in a mixture on the segment length** (log scale is used for the *x*-axis) for genome sets $M_{100}$ (circles) and $M_{28}$ (squars).

The curves presented in Figs. 5, 6 suggest that the fragments of length 40, 50 bp give better results than large-length fragments provided that the probability of losing a segment does not depend on its length. It should be noted that the results for almost collinear pairs of bacteria are qualitatively the same as we have already obtained with noise artificially introduced into the mixture vector. The results for the two most collinear pairs from set $M_{100}$ (see Table 1) are presented in Table SI7. The actual and the calculated multiplicities for each genome from set $M_{28}$ at C=50 and C=10000 are shown in Fig. 7.

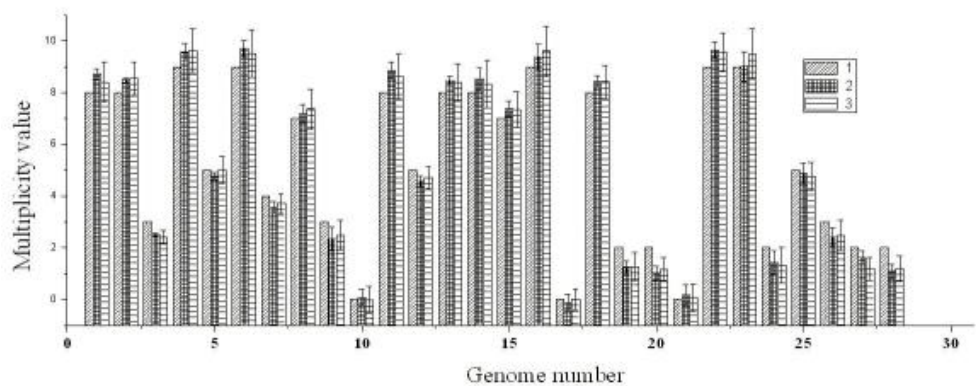

**Fig. 7. The actual (1) and the calculated multiplicities for each genome** from set $M_{28}$ at C=50 (2) and C=10000 (3).

## 5. Prospects for the extension of the model

Effect of the separating set growth. It has been shown above that certain violation of our basic model conditions, i.e., the assumption that the mixture genome set may not be a subset of the separating set (system (1) is inconsistent in this case), still allows applying the model quite effectively. In the cases analyzed above, the differences between these sets were minimal – the mixture set contained the genomes which did not belong to the separating set, but had almost collinear analogs there. In order to increase the probability of such a situation, the set of all sequenced genomes should be chosen as a separating set because we cannot, obviously, influence the composition of the mixture. Thus the efficiency of the method increases with the increase of the set of known genomes.

To illustrate this statement, in Fig.8, we show the dynamics of the angles between the new and the former sets of genomes over the last ten years[4]. It can be seen that in this period, these angles have been decreasing although each year, there appeared a genome significantly different from those sequenced before. Nevertheless, sooner or later, *the variety of microorganisms will be reduced to the variations of genomes around the forms already studied*. In this case, a mixture spectrum can be viewed as a sum of known genomic spectra and the same spectra with some variations. In other words, the spectra of unknown microorganisms will not differ significantly from those of the corresponding known microorganisms. Under these conditions, the multiplicities (coefficients) in the mixture of the known genomes can be obtained by the described-above method based on applying a bi-orthogonal basis or other methods of solving an inconsistent system. It is obvious (and we have shown it above) that the calculated multiplicities of genomes in the mixture are related not only to a particular genome, but also to all the other similar genomes, which, however, do not belong to the separating set (and thus are unknown). A plausible biological assumption is that these are unknown genomes which are close to this particular genome and encode similar biological traits. In this way, the qualitative contents of the mixture can be evaluated.

---

[4] Bacterial genomes and the year when they were sequenced were taken from the site
http://www.ncbi.nlm.nih.gov/genomes/lproks.cgi

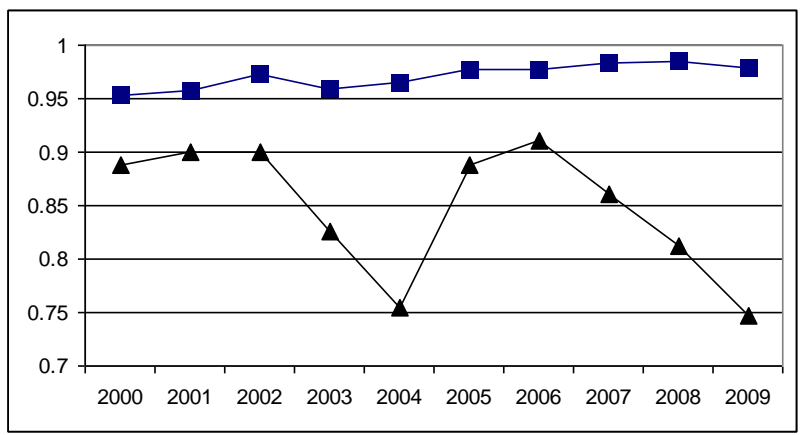

**Fig. 8. Dynamics of the angles between the new and the former sets of genomes over the last ten years**. X-axis: years. Y-axis: cosine values of the angles between CS of genomes. For each genome sequenced in a particular year, the minimal angle between this genome CS and CS of the genomes sequences up to this year is determined. The mean values of these angles cosines constitute the upper curve (squares). Each year, there appears a new genome which deviates from those already sequenced to the maximal extent, i.e. the one that has the greatest minimal angle. The lower curve (triangles) shows the cosines of these angles.

Linear genome space. Clearly, the expansion of the genome set requires an increase of the word space. For 6-letter words, the theoretically plausible limit of the space dimension is 4096 and the number of known genomes will soon exceed this value. Actually, the linear dimension of such a set is twice as small due to the existence of special word symmetry – extended *Chargaff*'s second parity rule [19]. This empirical rule, which claims that "reverse-complement" words (e.g. $ATTGC \Leftrightarrow GCAAT$) almost always have the same occurrence frequency in a genome.

Of course, it could be possible to work with words of larger length, e.g., 7, 8, or 10 bp. Obviously, the less is the length of the word chosen for constructing CS, the less may be the length of each fragment in the metagenome which this method is applied to. Additionally, it should be noted that bacterial genomes are usually of rather limited length and, therefore, relatively long words rarely occur in such genomes. For this reason, their occurrence frequencies become statistically unstable. For example, in a $10^6$ bp-long sequence, words 6, 7, 8, 9, and 10 bp in length occur, on average, 250, 62, 13, 3 times and only once, respectively.

Linear dimension that is generated by the set of 7- or 8-letter words will soon become less than the number of sequenced genomes. However, with regard to the described above extended Chargaff's rule, the linear dimension of the set of all 9-letter words is approximately 100,000. We suggest calculating each word's occurrence in the sequence even with one- or two-letter mismatch as it was done by us earlier [14]. Thus, along with each word, 351 words close to it (according to the standard evolutionary substitution metrics) also contribute to the total occurrence value. Such number of words ensures statistically significant occurrence values and the method has already proved to be effective, in particular, in the bacteria genome classification problems [20, 21]. An example of separating a genome mixture using a vocabulary that contains 200 10-letter words, the allowed mismatch being three letters, is shown in Fig. A1. Due to statistical stability, not all possible words of particular length have to be chosen as the basis; the number of such words is less and depends on the volume of the genome set under consideration.

## 6. Conclusion

In this work, the novel method of genome mixture separation proposed in [1] has been tested for separating a mixture that consists only of sequenced genomes. In the framework of the method, the formal solution of this problem presents no difficulty. However, it turned out that the conditionality of the problem is quite large, which requires estimating the solution quality depending on the data error (at least for medical purposes). We have evaluated the dependence of the solution quality on the fragment lengths in the metagenome, on random errors, etc. We have also proposed a method of adding a "neutral" bacterium to the metagenome, which allows estimating the impact of errors of different types on the solution quality in a real-life application of the method.

We believe that the method being studied extends far beyond the scope of the particular special mixture separation problem considered in this work. Indeed, it has been shown that the separation of a mixture is quite effective even if it contains not only the genomes that are known to be present in the mixture, but also their strains and even closely related bacteria genomes. It would be quite natural to suppose that with the increase of the number of sequenced genomes, an increasing number of unknown genomes will be located in a sufficiently close vicinity of the known ones, so that the method investigated in this work will become still more effective.

# Supporting Information for the article "Evaluation of the Genome Mixture Contents by means of the Compositional Spectra Method"

**Table SI1**. A set of 100 Eubacteria genomes, which represent all the main groups of bacteria. The number of genomes in each group is approximately proportional to the number of sequenced genomes in each group. The choice of genomes within the groups is random.

| N | Accession number | Name of bacterium/sequence |
|---|---|---|
| 1 | AE000657.1 | Aquifex aeolicus VF5, complete genome |
| 2 | AE001437.1 | Clostridium acetobutylicum ATCC 824, complete genome |
| 3 | AE003849.1 | Xylella fastidiosa 9a5c, complete genome |
| 4 | AE003852.1 | Vibrio cholerae O1 biovar eltor str. N16961 chromosome I, complete sequence |
| 5 | AE004439.1 | Pasteurella multocida subsp. multocida str. Pm70, complete genome |
| 6 | AE004969.1 | Neisseria gonorrhoeae FA 1090, complete genome |
| 7 | AE005176.1 | Lactococcus lactis subsp. lactis Il1403, complete genome |
| 8 | AE006468.1 | Salmonella enterica subsp. enterica serovar Typhimurium str. LT2 |
| 9 | AE007317.1 | Streptococcus pneumoniae R6, complete genome |
| 10 | AE007869.2 | Agrobacterium tumefaciens str. C58 circular chromosome, complete sequence |
| 11 | AE008691.1 | Thermoanaerobacter tengcongensis MB4, complete genome |
| 12 | AE008923.1 | Xanthomonas axonopodis pv. citri str. 306, complete genome |
| 13 | AE014291.4 | Brucella suis 1330 chromosome I, complete sequence |
| 14 | AE016830.1 | Enterococcus faecalis V583, complete genome |
| 15 | AE017125.1 | Helicobacter hepaticus ATCC 51449, complete genome |
| 16 | AE017143.1 | Haemophilus ducreyi strain 35000HP complete genome |
| 17 | AE017196.1 | Wolbachia endosymbiont of Drosophila melanogaster, complete genome |
| 18 | AE017262.2 | Listeria monocytogenes str. 4b F2365, complete genome |
| 19 | AE017334.2 | Bacillus anthracis str. 'Ames Ancestor', complete genome |
| 20 | AE017340.1 | Idiomarina loihiensis L2TR, complete genome |
| 21 | AE017354.1 | Legionella pneumophila subsp. pneumophila str. Philadelphia 1, complete genome |
| 22 | AJ938182.1 | Staphylococcus aureus RF122 complete genome |
| 23 | AL009126.3 | Bacillus subtilis subsp. subtilis str. 168 complete genome |
| 24 | AL590842.1 | Yersinia pestis CO92 complete genome |
| 25 | AL591824.1 | Listeria monocytogenes strain EGD, complete genome |
| 26 | AL645882.2 | Streptomyces coelicolor A3(2) complete genome |
| 27 | AL935263.1 | Lactobacillus plantarum strain WCFS1, complete genome |
| 28 | AM295007.1 | Streptococcus pyogenes Manfredo complete genome |
| 29 | AM412317.1 | Clostridium botulinum A str. ATCC 3502 complete genome |
| 30 | AM420293.1 | Saccharopolyspora erythraea NRRL2338 complete genome |
| 31 | AM884177.1 | Chlamydia trachomatis L2b/UCH-1/proctitis complete genome |
| 32 | AM889285.1 | Gluconacetobacter diazotrophicus PAl 5 complete genome |
| 33 | AP006618.1 | Nocardia farcinica IFM 10152 DNA, complete genome |
| 34 | AP006716.1 | Staphylococcus haemolyticus JCSC1435 DNA, complete genome |
| 35 | AP008971.1 | Finegoldia magna ATCC 29328 DNA, complete genome |
| 36 | AP010918.1 | Mycobacterium bovis BCG str. Tokyo 172 DNA, complete genome |
| 37 | BA000004.3 | Bacillus halodurans C-125 DNA, complete genome |
| 38 | BA000008.3 | Chlamydophila pneumoniae J138 genomic DNA, complete sequence |
| 39 | BA000012.4 | Mesorhizobium loti MAFF303099 DNA, complete genome |
| 40 | BA000022.2 | Synechocystis sp. PCC 6803 DNA, complete genome |
| 41 | BA000028.3 | Oceanobacillus iheyensis HTE831 DNA, complete genome |
| 42 | BA000034.2 | Streptococcus pyogenes SSI-1 DNA, complete genome |
| 43 | BA000039.2 | Thermosynechococcus elongatus BP-1 DNA, complete genome |
| 44 | BA000043.1 | Geobacillus kaustophilus HTA426 DNA, complete genome |
| 45 | BX293980.2 | Mycoplasma mycoides subsp. mycoides SC str. PG1, complete genome |
| 46 | BX470250.1 | Bordetella bronchiseptica strain RB50, complete genome |
| 47 | BX470251.1 | Photorhabdus luminescens subsp. laumondii TT01 complete genome |
| 48 | BX571857.1 | Staphylococcus aureus strain MSSA476, complete genome |
| 49 | BX897699.1 | Bartonella henselae strain Houston-1, complete genome |
| 50 | CP000011.1 | Burkholderia mallei ATCC 23344 chromosome 2, complete sequence |
| 51 | CP000025.1 | Campylobacter jejuni RM1221, complete genome |
| 52 | CP000033.3 | Lactobacillus acidophilus NCFM, complete genome |
| 53 | CP000050.1 | Xanthomonas campestris pv. campestris str. 8004, complete genome |
| 54 | CP000083.1 | Colwellia psychrerythraea 34H, complete genome |
| 55 | CP000094.2 | Pseudomonas fluorescens Pf0-1, complete genome |
| 56 | CP000139.1 | Bacteroides vulgatus ATCC 8482, complete genome |
| 57 | CP000141.1 | Carboxydothermus hydrogenoformans Z-2901, complete genome |
| 58 | CP000232.1 | Moorella thermoacetica ATCC 39073, complete genome |
| 59 | CP000246.1 | Clostridium perfringens ATCC 13124, complete genome |
| 60 | CP000282.1 | Saccharophagus degradans 2-40, complete genome |
| 61 | CP000285.1 | Chromohalobacter salexigens DSM 3043, complete genome |
| 62 | CP000359.1 | Deinococcus geothermalis DSM 11300, complete genome |
| 63 | CP000386.1 | Rubrobacter xylanophilus DSM 9941, complete genome |
| 64 | CP000390.1 | Mesorhizobium sp. BNC1, complete genome |
| 65 | CP000411.1 | Oenococcus oeni PSU-1, complete genome |
| 66 | CP000422.1 | Pediococcus pentosaceus ATCC 25745, complete genome |
| 67 | CP000423.1 | Lactobacillus casei ATCC 334, complete genome |
| 68 | CP000439.1 | Francisella tularensis subsp. novicida U112, complete genome |
| 69 | CP000478.1 | Syntrophobacter fumaroxidans MPOB, complete genome |
| 70 | CP000518.1 | Mycobacterium sp. KMS, complete genome |
| 71 | CP000527.1 | Desulfovibrio vulgaris DP4, complete genome |
| 72 | CP000612.1 | Desulfotomaculum reducens MI-1, complete genome |
| 73 | CP000698.1 | Geobacter uraniireducens Rf4, complete genome |
| 74 | CP000709.1 | Brucella ovis ATCC 25840 chromosome II, complete sequence |
| 75 | CP000717.1 | Mycobacterium tuberculosis F11, complete genome |
| 76 | CP000854.1 | Mycobacterium marinum M, complete genome |
| 77 | CP000909.1 | Chloroflexus aurantiacus J-10-fl, complete genome |
| 78 | CP000923.1 | Thermoanaerobacter sp. X514, complete genome |
| 79 | CP001022.1 | Exiguobacterium sibiricum 255-15, complete genome |
| 80 | CP001034.1 | Natranaerobius thermophilus JW/NM-WN-LF, complete genome |
| 81 | CP001037.1 | Nostoc punctiforme PCC 73102, complete genome |
| 82 | CP001098.1 | Halothermothrix orenii H 168, complete genome |
| 83 | CP001101.1 | Chlorobium phaeobacteroides BS1, complete genome |
| 84 | CP001154.1 | Laribacter hongkongensis HLHK9, complete genome |
| 85 | CP001213.1 | Bifidobacterium animalis subsp. lactis AD011, complete genome |
| 86 | CP001312.1 | Rhodobacter capsulatus SB 1003, complete genome |
| 87 | CP001472.1 | Acidobacterium capsulatum ATCC 51196, complete genome |
| 88 | CP001656.1 | Paenibacillus sp. JDR-2, complete genome |
| 89 | CP001785.1 | Ammonifex degensii KC4, complete genome |
| 90 | CP001999.1 | Arcobacter nitrofigilis DSM 7299, complete genome |
| 91 | CP002006.1 | Prevotella ruminicola 23, complete genome |
| 92 | CP002021.1 | Thiomonas intermedia K12, complete genome |
| 93 | CP002046.1 | Croceibacter atlanticus HTCC2559, complete genome |
| 94 | CP002056.1 | Methylotenera sp. 301, complete genome |
| 95 | CP002276.1 | Haemophilus influenzae R2846, complete genome |
| 96 | CP002277.1 | Haemophilus influenzae R2866, complete genome |
| 97 | CR543861.1 | Acinetobacter sp. ADP1 complete genome |
| 98 | CR626927.1 | Bacteroides fragilis NCTC 9343, complete genome |
| 99 | CR925677.1 | Ehrlichia ruminantium str. Gardel, complete genome |
| 100 | U00096.2 | Escherichia coli str. K-12 substr. MG1655, complete genome |

**Table SI2.** A set of 28 bacteria genomes, which are characterized in (9, SI, Fig. 9) as the most common gut bacteria.

| N | Accession number | Name of bacterium/sequence |
|---|---|---|
| 1 | NC_010278.1 | Actinobacillus pleuropneumoniae serovar 3 str. JL03, complete genome |
| 2 | NZ_DS264586.1 | Actinomyces odontolyticus ATCC 17982 Scfld021 genomic scaffold, whole genome |
| 3 | NC_009802.1 | Campylobacter concisus 13826, complete genome |
| 4 | NC_009792.1 | Citrobacter koseri ATCC BAA-895, complete genome |
| 5 | NC_008261.1 | Clostridium perfringens ATCC 13124, complete genome |
| 6 | NC_009778.1 | Cronobacter sakazakii ATCC BAA-894 chromosome, complete genome |
| 7 | NC_014618.1 | Enterobacter cloacae SCF1 chromosome, complete genome |
| 8 | CP000653.1 | Enterobacter sp. 638, complete genome |
| 9 | AC_000091.1 | Escherichia coli str. K-12 substr. W3110 strain K-12 |
| 10 | NC_011740.1 | Escherichia fergusonii ATCC 35469, complete genome |
| 11 | NC_007146.2 | Haemophilus influenzae 86-028NP, complete genome |
| 12 | NC_011852.1 | Haemophilus parasuis SH0165, complete genome |
| 13 | NC_010610.1 | Lactobacillus fermentum IFO 3956, complete genome |
| 14 | NC_008530.1 | Lactobacillus gasseri ATCC 33323, complete genome |
| 15 | NC_010080.1 | Lactobacillus helveticus DPC 4571, complete genome |
| 16 | NC_013504.1 | Lactobacillus johnsonii FI9785 chromosome, complete genome |
| 17 | NC_009513.1 | Lactobacillus reuteri DSM 20016, complete genome |
| 18 | NC_007576.1 | Lactobacillus sakei subsp. sakei 23K, complete genome |
| 19 | NC_007929.1 | Lactobacillus salivarius UCC118, complete genome |
| 20 | NC_002663.1 | Pasteurella multocida subsp. multocida str. Pm70, complete genome |
| 21 | NC_008525.1 | Pediococcus pentosaceus ATCC 25745, complete genome |
| 22 | NC_010067.1 | Salmonella enterica subsp. arizonae serovar 62:z4,z23:-- str. RSK2980 chromosome, |
| 23 | NC_009785.1 | Streptococcus gordonii str. Challis substr. CH1, complete genome |
| 24 | NC_013928.1 | Streptococcus mutans NN2025, complete genome |
| 25 | NC_014498.1 | Streptococcus pneumoniae 670-6B chromosome, complete genome |
| 26 | NC_002737.1 | Streptococcus pyogenes M1 GAS chromosome, complete genome |

| 27 | NC_009009.1 | Streptococcus sanguinis SK36, complete genome |
| 28 | NC_006449.1 | Streptococcus thermophilus CNRZ1066 chromosome, complete genome |

**Table SI3**. The results of the calculations of the genome multiplicities in the mixture of 100 genomes for different segment lengths for the deterministic case. **N** - genome number in Table SI1; **OM** - original multiplicity; 10, 20,..., 10000 - segment lengths in the mixture; the last column - mixture of whole genomes.

| N | OM | 10 | 20 | 30 | 40 | 50 | 100 | 200 | 500 | 1000 | 10000 | max |
|---|---|---|---|---|---|---|---|---|---|---|---|---|
| 1 | 8 | 7.91 | 7.96 | 7.87 | 7.93 | 8.01 | 7.96 | 7.97 | 8.02 | 8.02 | 8.01 | 8.00 |
| 2 | 8 | 7.13 | 7.69 | 7.45 | 7.92 | 8.00 | 8.01 | 7.94 | 7.96 | 7.96 | 7.95 | 8.00 |
| 3 | 3 | 3.25 | 2.65 | 3.26 | 2.84 | 3.01 | 2.85 | 2.96 | 2.97 | 3.00 | 2.99 | 3.00 |
| 4 | 9 | 8.87 | 9.12 | 9.06 | 8.78 | 8.92 | 9.11 | 8.96 | 8.97 | 8.95 | 8.91 | 9.00 |
| 5 | 5 | 5.90 | 5.32 | 5.32 | 5.07 | 5.57 | 5.29 | 5.09 | 5.08 | 5.03 | 5.09 | 5.00 |
| 6 | 9 | 8.83 | 8.99 | 9.06 | 8.94 | 9.05 | 9.03 | 8.98 | 9.01 | 8.98 | 8.99 | 9.00 |
| 7 | 4 | 3.69 | 3.96 | 3.74 | 3.77 | 3.60 | 3.86 | 3.88 | 3.85 | 3.97 | 4.09 | 4.00 |
| 8 | 7 | 7.75 | 7.00 | 7.31 | 6.78 | 7.12 | 7.14 | 7.02 | 6.94 | 6.94 | 6.95 | 7.00 |
| 9 | 3 | 3.77 | 2.89 | 2.96 | 3.44 | 3.32 | 3.13 | 3.18 | 3.10 | 3.08 | 3.10 | 3.00 |
| 10 | 0 | 0.61 | 0.12 | 0.02 | 0.13 | -0.01 | 0.08 | 0.05 | -0.07 | 0.01 | 0.00 | 0.00 |
| 11 | 8 | 9.42 | 8.56 | 8.22 | 8.11 | 8.37 | 8.20 | 7.97 | 8.03 | 7.95 | 7.95 | 8.00 |
| 12 | 5 | 5.84 | 5.60 | 4.81 | 5.27 | 4.78 | 4.82 | 4.94 | 4.89 | 4.90 | 5.02 | 5.00 |
| 13 | 8 | 8.03 | 8.23 | 6.97 | 7.78 | 7.71 | 7.64 | 7.86 | 7.86 | 7.86 | 7.92 | 8.00 |
| 14 | 8 | 7.97 | 8.13 | 7.90 | 8.09 | 7.82 | 7.94 | 7.95 | 7.95 | 7.98 | 7.97 | 8.00 |
| 15 | 7 | 7.39 | 7.13 | 7.09 | 7.10 | 7.09 | 7.07 | 7.04 | 7.01 | 7.00 | 7.00 | 7.00 |
| 16 | 9 | 8.02 | 8.87 | 8.63 | 8.94 | 8.65 | 8.91 | 8.94 | 8.98 | 9.02 | 9.04 | 9.00 |
| 17 | 0 | 0.09 | 0.14 | -0.04 | 0.22 | -0.01 | 0.09 | 0.09 | 0.03 | 0.03 | 0.04 | 0.00 |
| 18 | 8 | 9.04 | 7.13 | 7.58 | 7.41 | 8.28 | 7.77 | 7.95 | 7.95 | 7.89 | 7.91 | 8.00 |
| 19 | 2 | 2.03 | 1.97 | 1.47 | 2.02 | 2.30 | 2.23 | 2.13 | 2.11 | 2.05 | 1.92 | 2.00 |
| 20 | 2 | 2.62 | 2.47 | 2.54 | 2.06 | 2.06 | 2.16 | 2.08 | 2.16 | 2.08 | 1.98 | 2.00 |
| 21 | 0 | -0.28 | -0.26 | -0.25 | -0.18 | -0.10 | -0.08 | -0.11 | 0.00 | -0.01 | 0.02 | 0.00 |
| 22 | 9 | 11.79 | 10.47 | 8.88 | 9.87 | 8.91 | 9.19 | 9.01 | 9.16 | 9.03 | 8.76 | 9.00 |
| 23 | 9 | 9.06 | 9.09 | 9.11 | 9.23 | 9.00 | 8.84 | 8.99 | 8.95 | 9.02 | 8.98 | 9.00 |
| 24 | 2 | 2.95 | 2.09 | 1.85 | 1.94 | 2.24 | 1.99 | 2.06 | 2.00 | 2.06 | 2.13 | 2.00 |
| 25 | 5 | 4.29 | 5.51 | 5.39 | 5.32 | 4.68 | 5.16 | 5.10 | 5.06 | 5.07 | 5.12 | 5.00 |
| 26 | 3 | 2.62 | 2.63 | 2.98 | 2.81 | 2.98 | 2.92 | 2.91 | 2.97 | 2.98 | 3.00 | 3.00 |
| 27 | 2 | 1.09 | 1.82 | 1.53 | 2.09 | 2.03 | 1.94 | 1.99 | 2.00 | 2.00 | 1.99 | 2.00 |
| 28 | 2 | 1.24 | 1.51 | 1.40 | 1.69 | 1.59 | 1.67 | 1.86 | 2.03 | 2.01 | 1.93 | 1.99 |
| 29 | 1 | 1.13 | 1.16 | 1.25 | 1.05 | 1.39 | 1.24 | 1.14 | 1.11 | 1.02 | 1.04 | 1.00 |
| 30 | 6 | 5.87 | 6.08 | 6.01 | 6.03 | 6.06 | 6.02 | 6.05 | 6.00 | 6.02 | 5.96 | 6.00 |
| 31 | 5 | 4.71 | 4.70 | 4.80 | 4.83 | 4.89 | 4.95 | 4.96 | 5.03 | 5.00 | 5.02 | 5.00 |

| | | | | | | | | | | | |
|---|---|---|---|---|---|---|---|---|---|---|---|
| 32 | 8 | 7.73 | 7.94 | 7.94 | 7.91 | 7.97 | 7.95 | 7.98 | 8.05 | 8.04 | 7.94 | 8.00 |
| 33 | 0 | -0.07 | -0.27 | -0.15 | -0.13 | 0.08 | 0.07 | 0.08 | 0.01 | 0.02 | -0.03 | 0.00 |
| 34 | 2 | 2.55 | 2.16 | 2.07 | 1.95 | 2.04 | 1.90 | 1.94 | 1.94 | 1.96 | 1.87 | 2.00 |
| 35 | 6 | 5.94 | 6.07 | 5.95 | 5.90 | 5.94 | 5.95 | 5.96 | 5.99 | 5.99 | 5.97 | 6.00 |
| 36 | 4 | -3.10 | 2.85 | 3.09 | 3.43 | 4.96 | 4.51 | 4.70 | 3.64 | 3.49 | 4.13 | 3.99 |
| 37 | 2 | -0.05 | 1.32 | 1.90 | 1.61 | 1.70 | 1.97 | 2.01 | 2.00 | 2.02 | 2.05 | 2.00 |
| 38 | 8 | 8.63 | 8.38 | 8.19 | 8.30 | 8.02 | 7.97 | 7.99 | 7.95 | 7.98 | 8.04 | 8.00 |
| 39 | 3 | 2.17 | 2.49 | 2.90 | 2.72 | 2.94 | 2.84 | 2.88 | 2.95 | 2.94 | 3.00 | 3.00 |
| 40 | 9 | 9.28 | 9.20 | 9.25 | 8.88 | 9.15 | 9.08 | 9.06 | 8.95 | 8.95 | 8.90 | 9.00 |
| 41 | 3 | 3.53 | 3.34 | 3.52 | 2.99 | 3.19 | 3.02 | 2.86 | 2.99 | 3.01 | 3.01 | 3.00 |
| 42 | 7 | 8.12 | 7.73 | 7.95 | 7.27 | 7.64 | 7.45 | 7.19 | 6.97 | 6.98 | 6.89 | 7.00 |
| 43 | 8 | 7.62 | 7.86 | 7.80 | 7.93 | 7.95 | 8.09 | 7.99 | 8.06 | 8.06 | 8.09 | 8.00 |
| 44 | 6 | 5.69 | 5.82 | 5.95 | 5.89 | 5.91 | 5.89 | 5.93 | 6.04 | 6.04 | 5.97 | 6.00 |
| 45 | 3 | 3.09 | 2.99 | 3.15 | 2.99 | 3.03 | 3.04 | 3.02 | 3.03 | 3.03 | 3.01 | 3.00 |
| 46 | 5 | 5.35 | 5.26 | 4.93 | 5.25 | 5.37 | 5.27 | 5.14 | 5.02 | 5.03 | 5.01 | 5.00 |
| 47 | 6 | 5.51 | 6.13 | 6.02 | 5.91 | 6.19 | 6.18 | 6.09 | 6.01 | 5.91 | 6.03 | 6.00 |
| 48 | 5 | 3.50 | 4.07 | 5.24 | 4.48 | 5.01 | 4.87 | 4.90 | 4.85 | 4.90 | 5.14 | 5.00 |
| 49 | 0 | -0.21 | 0.48 | 0.00 | 0.37 | -0.05 | -0.10 | -0.04 | 0.01 | 0.05 | 0.10 | 0.00 |
| 50 | 8 | 7.99 | 7.97 | 8.04 | 7.98 | 7.97 | 7.95 | 8.00 | 7.96 | 8.00 | 8.05 | 8.00 |
| 51 | 0 | -0.47 | -0.32 | -0.12 | -0.21 | -0.15 | -0.12 | -0.05 | -0.07 | -0.05 | -0.03 | 0.00 |
| 52 | 1 | 0.97 | 0.71 | 1.15 | 0.95 | 0.96 | 0.88 | 0.99 | 1.02 | 0.99 | 1.05 | 1.00 |
| 53 | 3 | 1.99 | 2.38 | 3.26 | 2.85 | 3.08 | 3.23 | 3.10 | 3.21 | 3.12 | 2.97 | 3.00 |
| 54 | 4 | 3.79 | 3.50 | 4.06 | 3.75 | 4.31 | 3.95 | 3.96 | 3.97 | 3.98 | 4.07 | 4.00 |
| 55 | 4 | 4.26 | 4.22 | 4.21 | 4.06 | 4.02 | 4.13 | 4.05 | 3.95 | 3.97 | 4.03 | 4.00 |
| 56 | 8 | 8.20 | 7.82 | 8.14 | 8.07 | 8.07 | 7.97 | 8.07 | 7.97 | 8.03 | 8.10 | 8.00 |
| 57 | 0 | 0.56 | 0.36 | 0.00 | 0.23 | 0.18 | 0.12 | 0.01 | 0.01 | -0.02 | -0.02 | 0.00 |
| 58 | 8 | 8.34 | 8.25 | 8.04 | 8.08 | 7.96 | 8.05 | 8.05 | 8.07 | 8.08 | 8.11 | 8.00 |
| 59 | 2 | 2.61 | 2.22 | 2.18 | 2.21 | 1.76 | 1.80 | 2.11 | 1.91 | 2.05 | 2.08 | 2.00 |
| 60 | 0 | -0.03 | -0.01 | -0.12 | 0.18 | 0.05 | -0.06 | -0.02 | -0.03 | 0.00 | 0.00 | 0.00 |
| 61 | 7 | 6.65 | 6.78 | 6.86 | 6.96 | 6.99 | 6.95 | 6.94 | 7.02 | 7.02 | 6.96 | 7.00 |
| 62 | 0 | 0.02 | 0.08 | -0.17 | -0.02 | -0.06 | -0.01 | -0.02 | -0.04 | -0.05 | 0.04 | 0.00 |
| 63 | 1 | 1.28 | 1.25 | 1.15 | 1.15 | 1.02 | 1.01 | 1.00 | 1.01 | 0.99 | 1.01 | 1.00 |
| 64 | 6 | 7.45 | 6.90 | 6.54 | 6.33 | 6.09 | 6.26 | 6.18 | 6.10 | 6.08 | 6.06 | 6.00 |
| 65 | 2 | 2.33 | 2.21 | 2.10 | 2.10 | 2.41 | 2.24 | 2.09 | 2.11 | 2.02 | 1.98 | 2.00 |
| 66 | 8 | 8.62 | 8.30 | 8.00 | 8.09 | 7.90 | 8.05 | 8.08 | 8.01 | 8.00 | 8.00 | 8.00 |
| 67 | 2 | 2.86 | 2.38 | 2.07 | 2.13 | 1.73 | 1.93 | 1.99 | 2.00 | 2.04 | 1.95 | 2.00 |
| 68 | 2 | 1.02 | 1.70 | 1.97 | 1.91 | 1.57 | 1.72 | 1.80 | 1.97 | 1.98 | 1.99 | 2.00 |
| 69 | 5 | 4.92 | 4.53 | 4.86 | 4.73 | 4.93 | 5.01 | 4.96 | 4.98 | 4.96 | 5.00 | 5.00 |
| 70 | 6 | 6.42 | 6.38 | 5.84 | 6.33 | 5.96 | 6.10 | 6.07 | 6.13 | 6.03 | 5.99 | 6.00 |
| 71 | 6 | 6.04 | 6.14 | 5.94 | 6.07 | 5.93 | 6.05 | 6.02 | 6.02 | 6.00 | 5.96 | 6.00 |
| 72 | 1 | -0.26 | 0.34 | 0.92 | 0.47 | 0.73 | 0.71 | 0.78 | 0.87 | 0.92 | 1.06 | 1.00 |
| 73 | 9 | 9.06 | 8.87 | 8.93 | 9.09 | 9.04 | 8.94 | 9.08 | 9.00 | 9.02 | 8.96 | 9.00 |
| 74 | 3 | 2.62 | 2.42 | 3.57 | 3.04 | 3.28 | 3.16 | 3.03 | 3.15 | 3.06 | 3.05 | 3.00 |

| | | | | | | | | | | | |
|---|---|---|---|---|---|---|---|---|---|---|---|
| 75 | 3 | 9.81 | 4.62 | 3.89 | 3.71 | 1.89 | 2.55 | 2.18 | 3.31 | 3.41 | 2.71 3.00 |
| 76 | 0 | 0.46 | -0.20 | 0.43 | -0.07 | 0.20 | -0.13 | 0.09 | -0.04 | 0.05 | 0.10 0.00 |
| 77 | 8 | 7.71 | 7.95 | 8.07 | 7.98 | 8.00 | 8.01 | 7.97 | 7.97 | 7.98 | 7.98 8.00 |
| 78 | 4 | 3.45 | 3.65 | 4.38 | 3.90 | 3.73 | 3.91 | 4.05 | 4.01 | 4.05 | 4.05 4.00 |
| 79 | 4 | 4.05 | 3.94 | 4.04 | 4.04 | 4.03 | 4.05 | 3.98 | 4.00 | 3.98 | 4.03 4.00 |
| 80 | 9 | 8.27 | 8.75 | 8.76 | 8.90 | 9.09 | 9.05 | 8.91 | 9.06 | 9.05 | 8.97 9.00 |
| 81 | 2 | 2.02 | 1.76 | 2.14 | 2.12 | 1.97 | 1.89 | 2.02 | 1.96 | 1.97 | 2.01 2.00 |
| 82 | 5 | 5.14 | 4.84 | 4.95 | 5.16 | 4.74 | 4.84 | 4.98 | 4.93 | 4.92 | 4.95 5.00 |
| 83 | 0 | -0.04 | 0.06 | -0.06 | -0.09 | 0.01 | 0.00 | 0.01 | 0.01 | 0.00 | 0.02 0.00 |
| 84 | 8 | 8.01 | 8.03 | 8.00 | 7.95 | 8.01 | 7.95 | 7.98 | 7.99 | 8.00 | 7.96 8.00 |
| 85 | 4 | 3.79 | 3.95 | 4.02 | 3.94 | 3.95 | 3.91 | 3.92 | 3.96 | 3.98 | 3.99 4.00 |
| 86 | 5 | 4.90 | 5.17 | 5.12 | 5.10 | 4.91 | 4.98 | 4.98 | 4.97 | 4.99 | 5.03 5.00 |
| 87 | 3 | 2.48 | 2.95 | 2.42 | 2.94 | 3.16 | 3.02 | 3.05 | 3.05 | 3.03 | 3.04 3.00 |
| 88 | 1 | 0.88 | 1.17 | 1.07 | 1.09 | 0.92 | 1.08 | 1.03 | 0.99 | 0.97 | 0.96 1.00 |
| 89 | 7 | 6.54 | 6.73 | 7.04 | 6.84 | 6.95 | 6.96 | 6.96 | 6.98 | 6.99 | 6.94 7.00 |
| 90 | 2 | 2.06 | 2.44 | 1.97 | 2.04 | 2.09 | 2.12 | 2.05 | 2.02 | 1.98 | 1.99 2.00 |
| 91 | 2 | 2.27 | 2.08 | 2.20 | 1.97 | 1.97 | 2.07 | 2.01 | 2.00 | 1.98 | 1.98 2.00 |
| 92 | 3 | 3.04 | 2.90 | 3.16 | 2.89 | 3.00 | 2.93 | 2.91 | 3.01 | 3.00 | 2.99 3.00 |
| 93 | 4 | 3.87 | 3.83 | 3.84 | 3.79 | 4.01 | 3.99 | 3.93 | 4.03 | 4.02 | 4.04 4.00 |
| 94 | 7 | 7.31 | 7.45 | 7.09 | 7.17 | 7.12 | 7.19 | 7.10 | 6.92 | 6.99 | 7.13 7.00 |
| 95 | 1 | 0.48 | 0.83 | 1.57 | 0.93 | 1.52 | 1.47 | 1.21 | 1.02 | 0.99 | 0.93 1.00 |
| 96 | 4 | 4.07 | 3.68 | 3.18 | 4.05 | 3.01 | 3.27 | 3.73 | 4.00 | 4.03 | 3.98 4.00 |
| 97 | 8 | 8.08 | 8.07 | 7.95 | 8.11 | 8.09 | 8.02 | 8.08 | 8.01 | 8.01 | 7.96 8.00 |
| 98 | 5 | 4.68 | 5.12 | 5.06 | 5.14 | 4.85 | 4.98 | 5.04 | 4.95 | 5.01 | 4.96 5.00 |
| 99 | 7 | 7.27 | 7.09 | 6.96 | 6.95 | 7.07 | 7.13 | 7.07 | 7.02 | 7.02 | 6.98 7.00 |
| 100 | 2 | 0.20 | 1.22 | 1.44 | 2.01 | 1.62 | 1.69 | 1.89 | 2.03 | 2.06 | 2.01 2.00 |

**Table SI4**. The results of the calculations of the genome multiplicities in the mixture of 28 genomes for different segment lengths for the deterministic case. **N** - genome number in Table SI1; **OM** - original multiplicity; 10, 20,..., 10000 - segment lengths in the mixture; the last column - mixture of whole genomes.

| N | OM | 10 | 20 | 30 | 40 | 50 | 100 | 200 | 500 | 1000 | 10000 | max |
|---|---|---|---|---|---|---|---|---|---|---|---|---|
| 1 | 8 | 7.92 | 8.11 | 7.96 | 8.04 | 8.13 | 8.11 | 8.07 | 8.01 | 8.02 | 8.01 | 8.03 |
| 2 | 8 | 8.00 | 8.03 | 8.06 | 8.00 | 8.03 | 8.00 | 8.01 | 8.00 | 8.01 | 8.03 | 8.02 |
| 3 | 3 | 3.00 | 3.12 | 3.12 | 3.03 | 3.07 | 3.04 | 3.03 | 3.02 | 3.03 | 3.02 | 3.04 |
| 4 | 9 | 8.72 | 8.87 | 9.07 | 8.96 | 8.88 | 8.91 | 9.02 | 9.05 | 9.04 | 8.97 | 9.09 |
| 5 | 5 | 5.15 | 4.91 | 4.95 | 4.96 | 4.97 | 4.96 | 4.95 | 4.98 | 4.98 | 5.06 | 4.95 |
| 6 | 9 | 9.14 | 8.78 | 9.00 | 9.00 | 9.07 | 9.01 | 9.01 | 8.99 | 8.97 | 8.86 | 8.97 |
| 7 | 4 | 4.11 | 4.15 | 4.07 | 4.01 | 3.95 | 4.03 | 3.98 | 3.99 | 3.99 | 4.01 | 3.99 |
| 8 | 7 | 6.37 | 6.93 | 6.86 | 7.02 | 6.86 | 7.04 | 7.07 | 7.00 | 6.97 | 7.08 | 7.01 |
| 9 | 3 | 3.75 | 3.20 | 3.26 | 2.98 | 2.96 | 2.92 | 2.87 | 3.00 | 2.98 | 3.03 | 2.96 |
| 10 | 0 | -0.39 | -0.09 | -0.20 | 0.04 | 0.02 | -0.03 | 0.00 | -0.03 | 0.01 | 0.01 | -0.02 |
| 11 | 8 | 8.49 | 7.96 | 8.13 | 8.18 | 8.27 | 8.22 | 8.13 | 8.01 | 8.01 | 8.14 | 8.01 |

| | | | | | | | | | | | |
|---|---|---|---|---|---|---|---|---|---|---|---|
| 12 | 5 | 4.83 | 5.07 | 4.95 | 4.91 | 4.83 | 4.91 | 4.95 | 4.99 | 5.00 | 4.93 | 4.99 |
| 13 | 8 | 8.02 | 7.89 | 7.90 | 7.99 | 8.05 | 8.00 | 8.02 | 8.02 | 8.02 | 7.98 | 8.02 |
| 14 | 8 | 8.56 | 7.91 | 8.16 | 8.01 | 8.17 | 7.92 | 7.93 | 8.04 | 8.02 | 7.84 | 8.01 |
| 15 | 7 | 6.73 | 6.85 | 6.89 | 6.97 | 6.99 | 7.03 | 7.03 | 7.01 | 7.01 | 7.05 | 7.04 |
| 16 | 9 | 8.26 | 9.05 | 8.88 | 8.92 | 8.77 | 9.05 | 9.04 | 8.96 | 8.97 | 9.02 | 9.00 |
| 17 | 0 | -0.22 | 0.10 | 0.18 | -0.04 | -0.08 | -0.03 | -0.02 | 0.00 | -0.01 | 0.02 | -0.02 |
| 18 | 8 | 8.37 | 8.14 | 8.11 | 8.07 | 7.96 | 7.99 | 7.99 | 7.97 | 7.99 | 7.97 | 8.01 |
| 19 | 2 | 1.96 | 2.08 | 2.01 | 2.03 | 1.96 | 2.03 | 2.07 | 2.00 | 2.03 | 2.01 | 2.06 |
| 20 | 2 | 2.08 | 1.91 | 1.96 | 1.83 | 1.99 | 1.91 | 1.91 | 1.97 | 1.95 | 1.95 | 1.93 |
| 21 | 0 | -0.13 | 0.00 | 0.10 | 0.05 | 0.13 | 0.01 | -0.01 | 0.03 | 0.04 | 0.07 | -0.01 |
| 22 | 9 | 9.13 | 9.08 | 8.92 | 8.93 | 9.14 | 9.09 | 9.03 | 8.99 | 9.02 | 8.97 | 9.00 |
| 23 | 9 | 9.38 | 8.85 | 8.19 | 8.86 | 8.63 | 8.66 | 8.81 | 8.87 | 8.81 | 8.82 | 8.72 |
| 24 | 2 | 2.12 | 2.31 | 2.11 | 2.26 | 2.13 | 2.08 | 1.95 | 2.07 | 2.05 | 2.08 | 2.07 |
| 25 | 5 | 5.04 | 5.04 | 5.10 | 5.09 | 4.78 | 4.89 | 4.94 | 5.00 | 4.98 | 4.91 | 5.04 |
| 26 | 3 | 2.40 | 2.61 | 2.77 | 2.88 | 3.01 | 3.01 | 3.07 | 3.05 | 3.06 | 3.09 | 3.07 |
| 27 | 2 | 2.29 | 2.25 | 2.42 | 2.13 | 2.30 | 2.21 | 2.15 | 2.05 | 2.08 | 2.06 | 2.08 |
| 28 | 2 | 1.95 | 1.93 | 2.07 | 1.89 | 2.03 | 2.04 | 1.99 | 1.99 | 2.00 | 2.01 | 1.98 |

**Table SI5**. The mean multiplicity (*d*) and the squared deviation ($\sigma$) for each bacterium of set $M_{100}$. Averaging is performed over 100 experiments in each series. **N** - genome number in Table SI1; **OM** - original multiplicity; 10, 20,..., 10000 - segment lengths in the mixture. All the values are normalized by the 1$^{st}$ genome on the list.

| N | OM | 10 | | 20 | | 30 | | 40 | | 50 | | 100 | | 200 | | 500 | | 1000 | | 10000 | |
|---|---|---|---|---|---|---|---|---|---|---|---|---|---|---|---|---|---|---|---|---|---|
| 1 | 8 | 8.0 | 0.0 | 8.0 | 0.0 | 8.0 | 0.0 | 8.0 | 0.0 | 8.0 | 0.0 | 8.0 | 0.0 | 8.0 | 0.0 | 8.0 | 0.0 | 8.0 | 0.0 | 8.0 | 0.0 |
| 2 | 8 | 7.3 | 0.7 | 7.8 | 0.5 | 7.5 | 0.5 | 8.0 | 0.6 | 8.0 | 0.6 | 8.1 | 0.5 | 8.1 | 0.5 | 8.0 | 0.6 | 8.0 | 0.8 | 7.7 | 1.1 |
| 3 | 3 | 2.5 | 0.4 | 1.9 | 0.3 | 2.6 | 0.3 | 2.2 | 0.3 | 2.2 | 0.3 | 2.2 | 0.3 | 2.3 | 0.3 | 2.2 | 0.3 | 2.3 | 0.4 | 2.3 | 0.5 |
| 4 | 9 | 9.2 | 0.6 | 9.3 | 0.5 | 9.4 | 0.5 | 9.0 | 0.5 | 9.0 | 0.4 | 9.4 | 0.4 | 9.2 | 0.5 | 9.2 | 0.5 | 9.1 | 0.6 | 8.9 | 1.3 |
| 5 | 5 | 5.6 | 0.7 | 5.0 | 0.6 | 5.0 | 0.6 | 4.8 | 0.5 | 5.1 | 0.5 | 4.9 | 0.6 | 4.8 | 0.6 | 4.7 | 0.5 | 4.8 | 0.6 | 4.7 | 0.9 |
| 6 | 9 | 9.1 | 0.4 | 9.1 | 0.3 | 9.3 | 0.4 | 9.2 | 0.4 | 9.2 | 0.3 | 9.2 | 0.3 | 9.2 | 0.4 | 9.1 | 0.4 | 9.2 | 0.6 | 9.1 | 1.2 |
| 7 | 4 | 3.2 | 0.6 | 3.3 | 0.5 | 3.2 | 0.5 | 3.2 | 0.5 | 3.0 | 0.5 | 3.3 | 0.5 | 3.4 | 0.5 | 3.3 | 0.5 | 3.4 | 0.5 | 3.5 | 0.8 |
| 8 | 7 | 7.7 | 0.6 | 6.9 | 0.4 | 7.3 | 0.5 | 6.7 | 0.5 | 6.9 | 0.5 | 7.0 | 0.4 | 7.0 | 0.5 | 6.8 | 0.4 | 6.8 | 0.5 | 6.8 | 1.0 |
| 9 | 3 | 3.1 | 0.6 | 2.1 | 0.5 | 2.3 | 0.5 | 2.8 | 0.5 | 2.6 | 0.5 | 2.5 | 0.4 | 2.5 | 0.5 | 2.5 | 0.5 | 2.3 | 0.5 | 2.5 | 0.6 |
| 10 | 0 | 0.6 | 0.5 | 0.2 | 0.4 | 0.0 | 0.4 | 0.0 | 0.3 | 0.0 | 0.4 | 0.1 | 0.3 | 0.0 | 0.3 | -0.1 | 0.4 | 0.0 | 0.3 | -0.1 | 0.5 |
| 11 | 8 | 9.6 | 0.8 | 8.5 | 0.8 | 8.2 | 0.7 | 8.1 | 0.7 | 8.3 | 0.6 | 8.2 | 0.7 | 8.1 | 0.6 | 8.2 | 0.7 | 8.0 | 0.8 | 8.1 | 1.2 |
| 12 | 5 | 5.3 | 0.8 | 5.1 | 0.6 | 4.5 | 0.6 | 4.9 | 0.6 | 4.4 | 0.5 | 4.4 | 0.6 | 4.6 | 0.6 | 4.4 | 0.6 | 4.5 | 0.6 | 4.5 | 1.0 |
| 13 | 8 | 8.1 | 0.9 | 8.2 | 0.8 | 7.1 | 0.8 | 7.9 | 0.8 | 7.7 | 0.7 | 7.7 | 0.7 | 7.7 | 0.8 | 7.9 | 0.9 | 7.9 | 0.9 | 7.9 | 1.2 |
| 14 | 8 | 8.1 | 0.5 | 8.2 | 0.5 | 7.9 | 0.5 | 8.1 | 0.5 | 7.7 | 0.4 | 8.0 | 0.4 | 7.9 | 0.5 | 8.0 | 0.5 | 8.0 | 0.6 | 8.1 | 1.0 |
| 15 | 7 | 7.3 | 0.3 | 7.0 | 0.3 | 7.1 | 0.3 | 7.0 | 0.3 | 6.9 | 0.2 | 7.0 | 0.2 | 7.0 | 0.3 | 6.9 | 0.3 | 6.8 | 0.4 | 7.0 | 0.9 |
| 16 | 9 | 8.4 | 0.6 | 9.1 | 0.6 | 8.9 | 0.5 | 9.2 | 0.5 | 8.8 | 0.5 | 9.1 | 0.5 | 9.2 | 0.6 | 9.2 | 0.6 | 9.2 | 0.7 | 9.2 | 1.3 |
| 17 | 0 | 0.2 | 0.5 | 0.2 | 0.5 | 0.0 | 0.4 | 0.3 | 0.4 | 0.0 | 0.4 | 0.1 | 0.3 | 0.1 | 0.4 | 0.0 | 0.3 | 0.1 | 0.4 | 0.0 | 0.5 |
| 18 | 8 | 9.2 | 1.2 | 7.2 | 1.0 | 7.7 | 0.9 | 7.6 | 0.9 | 8.2 | 1.0 | 7.9 | 1.0 | 8.0 | 1.0 | 7.9 | 1.0 | 7.8 | 1.0 | 8.0 | 1.5 |
| 19 | 2 | 1.2 | 0.8 | 1.2 | 0.6 | 0.7 | 0.6 | 1.0 | 0.6 | 1.4 | 0.5 | 1.4 | 0.6 | 1.2 | 0.5 | 1.3 | 0.6 | 1.3 | 0.7 | 1.1 | 0.9 |
| 20 | 2 | 1.7 | 0.5 | 1.7 | 0.4 | 1.7 | 0.4 | 1.2 | 0.4 | 1.2 | 0.3 | 1.3 | 0.3 | 1.2 | 0.4 | 1.4 | 0.4 | 1.2 | 0.3 | 1.2 | 0.5 |
| 21 | 0 | -0.3 | 0.4 | -0.3 | 0.3 | -0.3 | 0.4 | -0.2 | 0.4 | -0.1 | 0.3 | -0.1 | 0.4 | -0.1 | 0.3 | 0.0 | 0.4 | 0.0 | 0.4 | 0.0 | 0.5 |
| 22 | 9 | 12.4 | 1.6 | 10.5 | 1.3 | 9.0 | 1.4 | 10.2 | 1.3 | 9.0 | 1.2 | 9.3 | 1.1 | 9.2 | 1.1 | 9.4 | 1.2 | 9.2 | 1.3 | 8.5 | 1.9 |
| 23 | 9 | 9.3 | 0.6 | 9.2 | 0.4 | 9.4 | 0.5 | 9.5 | 0.5 | 9.1 | 0.4 | 9.1 | 0.5 | 9.3 | 0.4 | 9.1 | 0.5 | 9.2 | 0.7 | 9.1 | 1.1 |
| 24 | 2 | 2.1 | 0.9 | 1.3 | 0.6 | 1.0 | 0.7 | 1.3 | 0.6 | 1.3 | 0.6 | 1.0 | 0.6 | 1.2 | 0.6 | 1.2 | 0.7 | 1.2 | 0.7 | 1.4 | 0.8 |

| | | | | | | | | | | | | | | | | | | | | |
|---|---|---|---|---|---|---|---|---|---|---|---|---|---|---|---|---|---|---|---|---|
| 25 | 5 | 3.8 | 0.9 | 5.0 | 0.9 | 5.0 | 0.7 | 4.9 | 0.7 | 4.3 | 0.7 | 4.7 | 0.7 | 4.8 | 0.7 | 4.7 | 0.8 | 4.7 | 0.8 | 4.6 | 1.0 |
| 26 | 3 | 1.9 | 0.3 | 1.9 | 0.2 | 2.3 | 0.2 | 2.1 | 0.2 | 2.2 | 0.2 | 2.2 | 0.2 | 2.2 | 0.2 | 2.2 | 0.2 | 2.3 | 0.2 | 2.4 | 0.4 |
| 27 | 2 | 0.3 | 0.5 | 0.9 | 0.4 | 0.6 | 0.4 | 1.2 | 0.4 | 1.2 | 0.4 | 1.1 | 0.4 | 1.2 | 0.4 | 1.1 | 0.4 | 1.1 | 0.4 | 1.2 | 0.5 |
| 28 | 2 | 0.4 | 1.5 | 0.8 | 1.2 | 0.4 | 1.1 | 0.7 | 1.0 | 0.8 | 1.0 | 0.8 | 1.2 | 1.1 | 1.2 | 1.1 | 1.2 | 1.2 | 1.2 | 0.9 | 1.8 |
| 29 | 1 | 0.1 | 0.6 | 0.1 | 0.5 | 0.2 | 0.5 | 0.1 | 0.4 | 0.3 | 0.4 | 0.2 | 0.4 | 0.1 | 0.4 | 0.2 | 0.5 | 0.0 | 0.5 | 0.0 | 0.5 |
| 30 | 6 | 5.7 | 0.6 | 5.8 | 0.5 | 5.9 | 0.5 | 5.9 | 0.5 | 5.9 | 0.4 | 5.8 | 0.5 | 5.9 | 0.4 | 5.8 | 0.4 | 5.8 | 0.5 | 5.7 | 1.0 |
| 31 | 5 | 4.4 | 0.5 | 4.3 | 0.4 | 4.4 | 0.4 | 4.4 | 0.4 | 4.4 | 0.4 | 4.5 | 0.4 | 4.6 | 0.4 | 4.7 | 0.5 | 4.6 | 0.5 | 4.6 | 0.7 |
| 32 | 8 | 7.8 | 0.4 | 7.9 | 0.4 | 8.0 | 0.3 | 8.0 | 0.3 | 7.9 | 0.3 | 8.0 | 0.3 | 8.1 | 0.3 | 8.1 | 0.5 | 8.1 | 0.5 | 8.0 | 1.1 |
| 33 | 0 | 0.0 | 0.3 | -0.3 | 0.3 | -0.2 | 0.3 | -0.1 | 0.3 | 0.0 | 0.3 | 0.0 | 0.3 | 0.1 | 0.3 | 0.0 | 0.3 | 0.0 | 0.3 | 0.1 | 0.4 |
| 34 | 2 | 1.7 | 0.6 | 1.4 | 0.5 | 1.3 | 0.5 | 1.1 | 0.5 | 1.2 | 0.4 | 1.0 | 0.5 | 1.1 | 0.5 | 1.1 | 0.5 | 1.2 | 0.5 | 1.1 | 0.7 |
| 35 | 6 | 5.8 | 0.4 | 5.9 | 0.4 | 5.7 | 0.3 | 5.6 | 0.3 | 5.6 | 0.4 | 5.7 | 0.3 | 5.7 | 0.4 | 5.7 | 0.4 | 5.7 | 0.5 | 5.7 | 0.9 |
| 36 | 4 | -3.0 | 4.5 | 2.7 | 3.8 | 2.7 | 3.6 | 2.9 | 4.1 | 4.7 | 3.2 | 4.4 | 3.7 | 4.1 | 3.3 | 3.3 | 3.9 | 2.4 | 4.0 | 3.1 | 4.4 |
| 37 | 2 | -0.9 | 0.7 | 0.4 | 0.6 | 1.2 | 0.5 | 0.8 | 0.5 | 0.9 | 0.5 | 1.0 | 0.6 | 1.1 | 0.6 | 1.1 | 0.6 | 1.1 | 0.6 | 1.1 | 0.8 |
| 38 | 8 | 8.7 | 0.6 | 8.4 | 0.5 | 8.3 | 0.5 | 8.4 | 0.5 | 8.0 | 0.5 | 8.1 | 0.5 | 8.1 | 0.5 | 7.9 | 0.6 | 8.0 | 0.7 | 8.1 | 1.2 |
| 39 | 3 | 1.5 | 0.4 | 1.8 | 0.3 | 2.2 | 0.3 | 2.1 | 0.3 | 2.2 | 0.3 | 2.1 | 0.3 | 2.2 | 0.3 | 2.2 | 0.3 | 2.2 | 0.3 | 2.2 | 0.5 |
| 40 | 9 | 9.5 | 0.5 | 9.4 | 0.4 | 9.5 | 0.5 | 9.1 | 0.4 | 9.3 | 0.4 | 9.2 | 0.4 | 9.3 | 0.4 | 9.1 | 0.4 | 9.2 | 0.5 | 9.0 | 1.3 |
| 41 | 3 | 2.7 | 0.6 | 2.5 | 0.5 | 2.8 | 0.6 | 2.3 | 0.5 | 2.4 | 0.5 | 2.4 | 0.5 | 2.1 | 0.5 | 2.2 | 0.5 | 2.3 | 0.5 | 2.2 | 0.7 |
| 42 | 7 | 7.9 | 1.4 | 7.5 | 1.2 | 8.0 | 1.0 | 7.3 | 1.0 | 7.4 | 1.0 | 7.3 | 1.1 | 7.1 | 1.1 | 6.9 | 1.1 | 6.9 | 1.3 | 6.9 | 2.1 |
| 43 | 8 | 7.7 | 0.5 | 7.8 | 0.4 | 7.9 | 0.4 | 8.0 | 0.4 | 7.9 | 0.4 | 8.1 | 0.5 | 8.0 | 0.4 | 8.1 | 0.5 | 8.1 | 0.6 | 8.1 | 1.2 |
| 44 | 6 | 5.5 | 0.4 | 5.5 | 0.3 | 5.8 | 0.3 | 5.7 | 0.3 | 5.6 | 0.3 | 5.7 | 0.3 | 5.7 | 0.3 | 5.8 | 0.3 | 5.8 | 0.4 | 5.8 | 0.8 |
| 45 | 3 | 2.4 | 0.2 | 2.3 | 0.2 | 2.4 | 0.2 | 2.3 | 0.2 | 2.3 | 0.2 | 2.4 | 0.2 | 2.3 | 0.2 | 2.3 | 0.2 | 2.3 | 0.2 | 2.3 | 0.3 |
| 46 | 5 | 4.9 | 0.6 | 4.9 | 0.5 | 4.7 | 0.5 | 4.8 | 0.4 | 4.9 | 0.4 | 4.9 | 0.4 | 4.8 | 0.4 | 4.6 | 0.5 | 4.7 | 0.5 | 4.5 | 0.8 |
| 47 | 6 | 5.3 | 0.9 | 5.8 | 0.7 | 5.9 | 0.7 | 5.6 | 0.7 | 6.0 | 0.7 | 6.0 | 0.6 | 5.9 | 0.7 | 5.7 | 0.8 | 5.6 | 0.7 | 5.8 | 1.1 |
| 48 | 5 | 3.0 | 1.1 | 3.7 | 0.8 | 5.0 | 0.9 | 4.0 | 0.9 | 4.5 | 0.9 | 4.6 | 0.8 | 4.6 | 0.8 | 4.4 | 0.9 | 4.4 | 0.8 | 5.0 | 1.1 |
| 49 | 0 | -0.3 | 0.6 | 0.5 | 0.5 | -0.1 | 0.6 | 0.4 | 0.5 | 0.0 | 0.5 | -0.2 | 0.6 | 0.0 | 0.5 | 0.1 | 0.5 | 0.1 | 0.6 | 0.2 | 0.9 |
| 50 | 8 | 8.1 | 0.3 | 8.0 | 0.3 | 8.1 | 0.3 | 8.0 | 0.3 | 8.0 | 0.3 | 8.0 | 0.2 | 8.0 | 0.3 | 8.0 | 0.3 | 8.0 | 0.5 | 8.0 | 1.0 |
| 51 | 0 | -0.4 | 0.3 | -0.3 | 0.2 | -0.1 | 0.2 | -0.2 | 0.2 | -0.2 | 0.2 | -0.1 | 0.2 | -0.1 | 0.2 | -0.1 | 0.2 | 0.0 | 0.2 | 0.0 | 0.3 |
| 52 | 1 | -0.1 | 0.3 | -0.3 | 0.3 | 0.1 | 0.3 | 0.0 | 0.3 | 0.0 | 0.3 | -0.1 | 0.3 | 0.0 | 0.3 | 0.0 | 0.3 | 0.0 | 0.3 | 0.1 | 0.4 |
| 53 | 3 | 1.5 | 0.9 | 1.7 | 0.7 | 2.5 | 0.7 | 2.2 | 0.7 | 2.3 | 0.6 | 2.5 | 0.7 | 2.3 | 0.7 | 2.5 | 0.7 | 2.4 | 0.7 | 2.4 | 1.1 |
| 54 | 4 | 3.3 | 0.7 | 2.9 | 0.6 | 3.5 | 0.6 | 3.3 | 0.5 | 3.6 | 0.5 | 3.4 | 0.5 | 3.4 | 0.5 | 3.4 | 0.6 | 3.4 | 0.6 | 3.4 | 0.8 |
| 55 | 4 | 3.8 | 0.4 | 3.6 | 0.3 | 3.7 | 0.4 | 3.5 | 0.3 | 3.5 | 0.3 | 3.5 | 0.3 | 3.5 | 0.3 | 3.4 | 0.3 | 3.4 | 0.4 | 3.5 | 0.6 |
| 56 | 8 | 8.2 | 0.5 | 7.8 | 0.4 | 8.3 | 0.4 | 8.1 | 0.4 | 8.1 | 0.4 | 8.0 | 0.4 | 8.1 | 0.4 | 8.0 | 0.5 | 8.1 | 0.6 | 8.2 | 1.2 |
| 57 | 0 | 0.6 | 0.4 | 0.3 | 0.3 | 0.0 | 0.3 | 0.2 | 0.2 | 0.2 | 0.3 | 0.1 | 0.2 | 0.0 | 0.3 | 0.1 | 0.3 | -0.1 | 0.3 | 0.0 | 0.4 |
| 58 | 8 | 8.4 | 0.5 | 8.3 | 0.4 | 8.2 | 0.4 | 8.2 | 0.4 | 8.0 | 0.3 | 8.1 | 0.4 | 8.1 | 0.4 | 8.1 | 0.4 | 8.1 | 0.6 | 8.0 | 1.2 |
| 59 | 2 | 1.7 | 0.6 | 1.4 | 0.6 | 1.3 | 0.6 | 1.4 | 0.5 | 0.9 | 0.5 | 1.0 | 0.6 | 1.3 | 0.6 | 1.0 | 0.5 | 1.3 | 0.6 | 1.3 | 0.8 |
| 60 | 0 | -0.1 | 0.3 | 0.0 | 0.2 | -0.1 | 0.3 | 0.2 | 0.2 | 0.0 | 0.2 | 0.0 | 0.2 | 0.0 | 0.2 | 0.0 | 0.2 | 0.0 | 0.2 | 0.0 | 0.3 |
| 61 | 7 | 6.6 | 0.4 | 6.6 | 0.3 | 6.8 | 0.3 | 6.8 | 0.3 | 6.8 | 0.3 | 6.8 | 0.3 | 6.9 | 0.4 | 6.8 | 0.4 | 7.0 | 0.5 | 6.8 | 1.0 |
| 62 | 0 | 0.1 | 0.3 | 0.1 | 0.3 | -0.1 | 0.2 | 0.0 | 0.3 | -0.1 | 0.2 | 0.0 | 0.3 | 0.0 | 0.2 | 0.0 | 0.3 | -0.1 | 0.3 | 0.0 | 0.4 |
| 63 | 1 | 0.3 | 0.2 | 0.3 | 0.2 | 0.1 | 0.2 | 0.1 | 0.1 | 0.0 | 0.1 | 0.0 | 0.1 | 0.0 | 0.1 | 0.0 | 0.1 | 0.0 | 0.2 | 0.0 | 0.2 |
| 64 | 6 | 7.2 | 0.8 | 6.7 | 0.6 | 6.3 | 0.6 | 6.1 | 0.6 | 5.7 | 0.6 | 6.1 | 0.6 | 5.9 | 0.6 | 5.9 | 0.6 | 5.7 | 0.7 | 5.7 | 1.1 |
| 65 | 2 | 1.5 | 0.3 | 1.4 | 0.3 | 1.3 | 0.3 | 1.3 | 0.2 | 1.6 | 0.3 | 1.4 | 0.3 | 1.3 | 0.3 | 1.2 | 0.3 | 1.2 | 0.3 | 1.1 | 0.4 |
| 66 | 8 | 8.8 | 0.6 | 8.3 | 0.6 | 8.1 | 0.6 | 8.1 | 0.6 | 7.9 | 0.6 | 8.1 | 0.5 | 8.1 | 0.5 | 8.0 | 0.6 | 8.1 | 0.8 | 8.1 | 1.3 |
| 67 | 2 | 1.9 | 0.4 | 1.5 | 0.4 | 1.3 | 0.5 | 1.3 | 0.4 | 0.9 | 0.4 | 1.2 | 0.4 | 1.1 | 0.4 | 1.2 | 0.4 | 1.2 | 0.4 | 1.1 | 0.5 |
| 68 | 2 | 0.2 | 0.5 | 0.9 | 0.4 | 1.1 | 0.3 | 1.0 | 0.3 | 0.8 | 0.3 | 0.9 | 0.3 | 0.9 | 0.3 | 1.1 | 0.3 | 1.2 | 0.3 | 1.1 | 0.4 |
| 69 | 5 | 4.6 | 0.5 | 4.1 | 0.5 | 4.6 | 0.5 | 4.3 | 0.4 | 4.5 | 0.4 | 4.6 | 0.3 | 4.6 | 0.4 | 4.6 | 0.4 | 4.5 | 0.5 | 4.6 | 0.8 |
| 70 | 6 | 6.1 | 0.6 | 6.2 | 0.5 | 5.7 | 0.5 | 6.1 | 0.5 | 5.7 | 0.4 | 6.0 | 0.5 | 5.8 | 0.5 | 5.8 | 0.5 | 5.8 | 0.6 | 5.6 | 0.9 |
| 71 | 6 | 5.9 | 0.3 | 5.9 | 0.3 | 5.7 | 0.3 | 5.9 | 0.3 | 5.6 | 0.3 | 5.8 | 0.3 | 5.8 | 0.3 | 5.8 | 0.3 | 5.7 | 0.4 | 5.7 | 0.8 |
| 72 | 1 | -1.2 | 0.6 | -0.7 | 0.5 | 0.0 | 0.5 | -0.6 | 0.5 | -0.3 | 0.5 | -0.2 | 0.6 | -0.2 | 0.6 | -0.2 | 0.5 | -0.2 | 0.6 | 0.0 | 0.7 |
| 73 | 9 | 9.3 | 0.6 | 9.0 | 0.5 | 9.1 | 0.5 | 9.3 | 0.4 | 9.1 | 0.4 | 9.2 | 0.5 | 9.3 | 0.5 | 9.2 | 0.4 | 9.2 | 0.6 | 9.2 | 1.4 |
| 74 | 3 | 2.0 | 0.6 | 1.7 | 0.5 | 2.9 | 0.5 | 2.3 | 0.5 | 2.6 | 0.5 | 2.5 | 0.5 | 2.4 | 0.4 | 2.4 | 0.5 | 2.4 | 0.5 | 2.4 | 0.7 |
| 75 | 3 | 8.7 | 4.2 | 3.6 | 3.7 | 3.1 | 3.4 | 3.0 | 4.0 | 0.8 | 3.0 | 1.3 | 3.7 | 1.5 | 3.2 | 2.4 | 3.7 | 3.4 | 3.7 | 2.5 | 4.0 |

| N | | 10 | | 20 | | 30 | | 40 | | 50 | | 100 | | 200 | | 500 | | 1000 | | 10000 | |
|---|---|---|---|---|---|---|---|---|---|---|---|---|---|---|---|---|---|---|---|---|---|
| 76 | 0 | 0.4 | 0.8 | -0.2 | 0.6 | 0.3 | 0.6 | -0.1 | 0.6 | 0.3 | 0.6 | 0.0 | 0.6 | 0.1 | 0.6 | -0.1 | 0.5 | 0.0 | 0.7 | 0.1 | 0.9 |
| 77 | 8 | 7.8 | 0.4 | 8.0 | 0.3 | 8.2 | 0.3 | 8.0 | 0.3 | 8.0 | 0.3 | 8.0 | 0.3 | 8.1 | 0.3 | 8.0 | 0.3 | 8.0 | 0.5 | 8.1 | 0.9 |
| 78 | 4 | 2.8 | 0.6 | 3.1 | 0.6 | 3.9 | 0.5 | 3.3 | 0.5 | 3.2 | 0.5 | 3.4 | 0.5 | 3.5 | 0.5 | 3.3 | 0.5 | 3.5 | 0.5 | 3.5 | 0.8 |
| 79 | 4 | 3.5 | 0.4 | 3.4 | 0.3 | 3.5 | 0.3 | 3.5 | 0.3 | 3.4 | 0.3 | 3.5 | 0.3 | 3.4 | 0.3 | 3.4 | 0.3 | 3.4 | 0.4 | 3.5 | 0.5 |
| 80 | 9 | 8.5 | 0.5 | 8.9 | 0.5 | 9.0 | 0.5 | 9.1 | 0.5 | 9.2 | 0.4 | 9.3 | 0.5 | 9.2 | 0.5 | 9.2 | 0.5 | 9.1 | 0.6 | 9.1 | 1.2 |
| 81 | 2 | 1.2 | 0.5 | 1.0 | 0.4 | 1.3 | 0.4 | 1.3 | 0.3 | 1.2 | 0.4 | 1.0 | 0.4 | 1.2 | 0.3 | 1.1 | 0.4 | 1.1 | 0.4 | 1.2 | 0.5 |
| 82 | 5 | 4.8 | 0.5 | 4.4 | 0.4 | 4.6 | 0.4 | 4.8 | 0.4 | 4.3 | 0.4 | 4.4 | 0.4 | 4.6 | 0.4 | 4.5 | 0.4 | 4.6 | 0.5 | 4.6 | 0.8 |
| 83 | 0 | -0.1 | 0.3 | 0.0 | 0.3 | 0.0 | 0.3 | -0.1 | 0.3 | 0.0 | 0.3 | 0.0 | 0.3 | 0.0 | 0.3 | 0.1 | 0.3 | 0.0 | 0.3 | 0.0 | 0.4 |
| 84 | 8 | 8.0 | 0.5 | 8.1 | 0.3 | 8.1 | 0.3 | 8.0 | 0.4 | 8.0 | 0.3 | 8.0 | 0.3 | 8.0 | 0.4 | 8.0 | 0.4 | 8.0 | 0.5 | 8.0 | 1.1 |
| 85 | 4 | 3.2 | 0.3 | 3.4 | 0.3 | 3.5 | 0.2 | 3.4 | 0.2 | 3.4 | 0.2 | 3.4 | 0.2 | 3.4 | 0.2 | 3.4 | 0.3 | 3.4 | 0.3 | 3.4 | 0.5 |
| 86 | 5 | 4.5 | 0.3 | 4.7 | 0.2 | 4.7 | 0.2 | 4.7 | 0.2 | 4.5 | 0.2 | 4.6 | 0.2 | 4.6 | 0.2 | 4.5 | 0.3 | 4.6 | 0.3 | 4.6 | 0.7 |
| 87 | 3 | 1.9 | 0.5 | 2.2 | 0.4 | 1.8 | 0.4 | 2.2 | 0.4 | 2.5 | 0.4 | 2.3 | 0.4 | 2.4 | 0.3 | 2.3 | 0.4 | 2.3 | 0.4 | 2.3 | 0.6 |
| 88 | 1 | -0.1 | 0.2 | 0.1 | 0.3 | 0.0 | 0.3 | 0.1 | 0.2 | -0.1 | 0.2 | 0.1 | 0.3 | 0.0 | 0.3 | 0.0 | 0.2 | 0.0 | 0.3 | 0.0 | 0.3 |
| 89 | 7 | 6.5 | 0.4 | 6.6 | 0.4 | 7.0 | 0.3 | 6.7 | 0.3 | 6.8 | 0.3 | 6.8 | 0.3 | 6.9 | 0.4 | 6.8 | 0.4 | 6.9 | 0.6 | 6.8 | 1.0 |
| 90 | 2 | 1.3 | 0.4 | 1.6 | 0.4 | 1.2 | 0.4 | 1.2 | 0.3 | 1.2 | 0.3 | 1.2 | 0.3 | 1.2 | 0.3 | 1.1 | 0.3 | 1.1 | 0.4 | 1.1 | 0.4 |
| 91 | 2 | 1.4 | 0.3 | 1.2 | 0.3 | 1.4 | 0.4 | 1.2 | 0.3 | 1.1 | 0.3 | 1.2 | 0.3 | 1.1 | 0.4 | 1.1 | 0.3 | 1.1 | 0.4 | 1.2 | 0.5 |
| 92 | 3 | 2.3 | 0.3 | 2.2 | 0.3 | 2.4 | 0.3 | 2.2 | 0.3 | 2.3 | 0.3 | 2.3 | 0.3 | 2.2 | 0.3 | 2.3 | 0.3 | 2.3 | 0.3 | 2.3 | 0.5 |
| 93 | 4 | 3.3 | 0.3 | 3.3 | 0.2 | 3.3 | 0.2 | 3.3 | 0.2 | 3.4 | 0.2 | 3.4 | 0.2 | 3.4 | 0.2 | 3.4 | 0.3 | 3.4 | 0.3 | 3.5 | 0.5 |
| 94 | 7 | 7.2 | 0.7 | 7.3 | 0.6 | 7.1 | 0.6 | 7.0 | 0.6 | 7.0 | 0.6 | 7.0 | 0.5 | 7.0 | 0.6 | 6.8 | 0.6 | 6.9 | 0.6 | 6.9 | 1.1 |
| 95 | 1 | -0.5 | 1.0 | -0.4 | 1.0 | 0.5 | 0.8 | 0.0 | 0.8 | 0.6 | 0.7 | 0.4 | 0.9 | 0.1 | 0.7 | 0.0 | 0.9 | -0.1 | 0.9 | -0.1 | 0.8 |
| 96 | 4 | 3.4 | 1.3 | 3.3 | 1.1 | 2.7 | 1.0 | 3.4 | 1.0 | 2.4 | 0.9 | 2.7 | 1.0 | 3.3 | 0.9 | 3.4 | 1.1 | 3.6 | 1.1 | 3.5 | 1.1 |
| 97 | 8 | 8.2 | 0.5 | 8.1 | 0.4 | 8.0 | 0.4 | 8.1 | 0.4 | 8.0 | 0.4 | 8.1 | 0.4 | 8.2 | 0.4 | 8.0 | 0.4 | 8.0 | 0.6 | 8.0 | 1.1 |
| 98 | 5 | 4.4 | 0.5 | 4.7 | 0.4 | 4.7 | 0.4 | 4.8 | 0.4 | 4.4 | 0.3 | 4.6 | 0.3 | 4.7 | 0.4 | 4.5 | 0.4 | 4.6 | 0.4 | 4.5 | 0.8 |
| 99 | 7 | 7.1 | 0.3 | 6.9 | 0.3 | 6.9 | 0.4 | 6.8 | 0.3 | 6.9 | 0.3 | 7.0 | 0.3 | 7.0 | 0.3 | 6.9 | 0.4 | 6.9 | 0.5 | 6.9 | 1.1 |
| 100 | 2 | -0.8 | 0.6 | 0.3 | 0.6 | 0.6 | 0.5 | 1.1 | 0.5 | 0.8 | 0.5 | 1.0 | 0.5 | 1.1 | 0.5 | 1.2 | 0.5 | 1.2 | 0.5 | 1.2 | 0.7 |

**Table SI6**. The mean multiplicity (*d*) and the squared deviation ($\sigma$) for each bacterium of set $M_{28}$. Averaging is performed over 100 experiments in each series. **N** - genome number in Table SI1; **OM** - original multiplicity; 10, 20, ..., 10000 - segment lengths in the mixture. All the values are normalized by the 1$^{st}$ genome on the list.

| N | OM | 10 | | 20 | | 30 | | 40 | | 50 | | 100 | | 200 | | 500 | | 1000 | | 10000 | |
|---|---|---|---|---|---|---|---|---|---|---|---|---|---|---|---|---|---|---|---|---|---|
| 1 | 8 | 8.0 | 0.0 | 8.0 | 0.0 | 8.0 | 0.0 | 8.0 | 0.0 | 8.0 | 0.0 | 8.0 | 0.0 | 8.0 | 0.0 | 8.0 | 0.0 | 8.0 | 0.0 | 8.0 | 0.0 |
| 2 | 8 | 8.0 | 0.2 | 7.9 | 0.2 | 8.1 | 0.2 | 8.0 | 0.2 | 7.9 | 0.2 | 7.9 | 0.2 | 7.9 | 0.2 | 8.0 | 0.3 | 7.9 | 0.4 | 8.2 | 1.0 |
| 3 | 3 | 2.3 | 0.1 | 2.4 | 0.1 | 2.4 | 0.1 | 2.3 | 0.1 | 2.3 | 0.1 | 2.3 | 0.1 | 2.3 | 0.1 | 2.3 | 0.1 | 2.3 | 0.1 | 2.3 | 0.3 |
| 4 | 9 | 8.9 | 0.4 | 8.8 | 0.3 | 9.2 | 0.4 | 9.0 | 0.4 | 8.8 | 0.3 | 9.0 | 0.3 | 9.1 | 0.4 | 9.1 | 0.4 | 9.1 | 0.6 | 9.3 | 1.3 |
| 5 | 5 | 4.7 | 0.2 | 4.4 | 0.2 | 4.5 | 0.2 | 4.6 | 0.2 | 4.5 | 0.2 | 4.5 | 0.2 | 4.5 | 0.2 | 4.6 | 0.3 | 4.5 | 0.3 | 4.8 | 0.7 |
| 6 | 9 | 9.3 | 0.4 | 8.8 | 0.4 | 9.2 | 0.4 | 9.1 | 0.4 | 9.1 | 0.3 | 9.1 | 0.4 | 9.0 | 0.4 | 9.2 | 0.5 | 9.0 | 0.6 | 9.1 | 1.2 |
| 7 | 4 | 3.6 | 0.3 | 3.5 | 0.3 | 3.5 | 0.2 | 3.4 | 0.3 | 3.4 | 0.2 | 3.4 | 0.2 | 3.4 | 0.2 | 3.5 | 0.3 | 3.4 | 0.3 | 3.6 | 0.5 |
| 8 | 7 | 6.2 | 0.4 | 6.7 | 0.4 | 6.7 | 0.3 | 6.9 | 0.3 | 6.6 | 0.4 | 6.8 | 0.4 | 6.9 | 0.4 | 6.9 | 0.4 | 6.7 | 0.5 | 7.1 | 1.0 |
| 9 | 3 | 3.1 | 0.6 | 2.5 | 0.4 | 2.6 | 0.5 | 2.3 | 0.4 | 2.2 | 0.4 | 2.2 | 0.4 | 2.1 | 0.5 | 2.3 | 0.5 | 2.2 | 0.5 | 2.4 | 0.6 |
| 10 | 0 | -0.4 | 0.4 | -0.1 | 0.3 | -0.2 | 0.3 | 0.0 | 0.3 | 0.0 | 0.3 | 0.0 | 0.3 | 0.0 | 0.3 | -0.1 | 0.3 | 0.0 | 0.4 | 0.0 | 0.5 |
| 11 | 8 | 8.5 | 0.4 | 7.9 | 0.3 | 8.2 | 0.4 | 8.2 | 0.3 | 8.1 | 0.3 | 8.2 | 0.4 | 8.1 | 0.4 | 8.1 | 0.5 | 7.9 | 0.5 | 8.3 | 1.4 |
| 12 | 5 | 4.5 | 0.3 | 4.6 | 0.2 | 4.6 | 0.3 | 4.5 | 0.2 | 4.3 | 0.2 | 4.4 | 0.2 | 4.5 | 0.2 | 4.6 | 0.3 | 4.5 | 0.3 | 4.5 | 0.6 |
| 13 | 8 | 8.1 | 0.3 | 7.8 | 0.2 | 7.9 | 0.2 | 8.0 | 0.2 | 7.9 | 0.2 | 7.9 | 0.3 | 8.0 | 0.2 | 8.0 | 0.3 | 8.0 | 0.4 | 8.1 | 1.1 |
| 14 | 8 | 8.6 | 0.6 | 7.8 | 0.4 | 8.1 | 0.4 | 8.0 | 0.4 | 8.0 | 0.4 | 7.8 | 0.4 | 7.9 | 0.4 | 8.0 | 0.5 | 7.9 | 0.5 | 8.0 | 1.1 |
| 15 | 7 | 6.6 | 0.3 | 6.6 | 0.3 | 6.8 | 0.3 | 6.8 | 0.3 | 6.8 | 0.3 | 6.8 | 0.3 | 6.8 | 0.3 | 6.9 | 0.4 | 6.8 | 0.4 | 7.1 | 0.9 |
| 16 | 9 | 8.5 | 0.6 | 9.0 | 0.5 | 9.1 | 0.5 | 9.0 | 0.5 | 8.8 | 0.5 | 9.1 | 0.5 | 9.1 | 0.5 | 9.1 | 0.5 | 9.0 | 0.7 | 9.2 | 1.4 |
| 17 | 0 | -0.3 | 0.4 | 0.0 | 0.4 | 0.2 | 0.3 | 0.0 | 0.3 | -0.1 | 0.3 | 0.0 | 0.3 | 0.0 | 0.3 | 0.0 | 0.3 | 0.0 | 0.3 | 0.0 | 0.4 |
| 18 | 8 | 8.4 | 0.4 | 8.0 | 0.3 | 8.2 | 0.4 | 8.0 | 0.3 | 7.8 | 0.3 | 7.9 | 0.3 | 8.0 | 0.3 | 8.0 | 0.4 | 7.9 | 0.4 | 8.1 | 1.0 |
| 19 | 2 | 1.1 | 0.3 | 1.2 | 0.3 | 1.2 | 0.2 | 1.1 | 0.2 | 1.1 | 0.2 | 1.2 | 0.2 | 1.2 | 0.2 | 1.1 | 0.3 | 1.2 | 0.3 | 1.2 | 0.5 |
| 20 | 2 | 1.1 | 0.3 | 1.0 | 0.3 | 1.1 | 0.3 | 1.0 | 0.2 | 1.1 | 0.3 | 1.0 | 0.3 | 1.1 | 0.3 | 1.1 | 0.3 | 1.1 | 0.3 | 1.1 | 0.5 |
| 21 | 0 | -0.1 | 0.5 | 0.0 | 0.3 | 0.0 | 0.4 | 0.1 | 0.3 | 0.1 | 0.4 | 0.0 | 0.4 | 0.0 | 0.4 | 0.0 | 0.4 | -0.1 | 0.4 | 0.1 | 0.5 |
| 22 | 9 | 9.3 | 0.5 | 9.1 | 0.4 | 9.1 | 0.4 | 9.1 | 0.4 | 9.1 | 0.4 | 9.1 | 0.4 | 9.0 | 0.4 | 9.2 | 0.5 | 9.0 | 0.6 | 9.2 | 1.2 |
| 23 | 9 | 9.6 | 0.7 | 8.9 | 0.6 | 8.4 | 0.5 | 8.9 | 0.5 | 8.6 | 0.5 | 8.7 | 0.5 | 8.9 | 0.7 | 9.0 | 0.6 | 8.8 | 0.6 | 9.1 | 1.4 |
| 24 | 2 | 1.3 | 0.4 | 1.5 | 0.4 | 1.3 | 0.4 | 1.3 | 0.4 | 1.2 | 0.4 | 1.2 | 0.4 | 1.1 | 0.4 | 1.3 | 0.5 | 1.1 | 0.4 | 1.3 | 0.7 |
| 25 | 5 | 4.6 | 0.4 | 4.5 | 0.4 | 4.7 | 0.4 | 4.6 | 0.4 | 4.3 | 0.4 | 4.5 | 0.3 | 4.4 | 0.4 | 4.6 | 0.4 | 4.4 | 0.4 | 4.6 | 0.6 |
| 26 | 3 | 1.7 | 0.5 | 1.9 | 0.3 | 2.0 | 0.4 | 2.2 | 0.3 | 2.3 | 0.3 | 2.3 | 0.4 | 2.3 | 0.4 | 2.3 | 0.4 | 2.3 | 0.4 | 2.4 | 0.6 |
| 27 | 2 | 1.4 | 0.4 | 1.3 | 0.3 | 1.6 | 0.3 | 1.3 | 0.2 | 1.4 | 0.2 | 1.3 | 0.3 | 1.3 | 0.3 | 1.2 | 0.3 | 1.2 | 0.3 | 1.1 | 0.4 |
| 28 | 2 | 1.1 | 0.3 | 1.0 | 0.3 | 1.2 | 0.3 | 1.1 | 0.2 | 1.2 | 0.3 | 1.1 | 0.3 | 1.1 | 0.3 | 1.1 | 0.3 | 1.2 | 0.3 | 1.1 | 0.5 |

| N | AM | 10 | 20 | 30 | 40 | 50 |
|---|---|---|---|---|---|---|
| 36 | 4 | -3.01 | 2.67 | 3.68 | 2.86 | 4.68 |
| 75 | 3 | 8.74 | 3.58 | 3.11 | 3 | 0.77 |
| sum | 7 | 5.73 | 6.25 | 6.79 | 5.86 | 5.45 |
| | | | | | | |
| 28 | 2 | 0.43 | 0.77 | 0.43 | 0.69 | 0.82 |
| 42 | 7 | 7.92 | 7.52 | 8 | 7.28 | 7.37 |
| Sum | 9 | 8.35 | 8.29 | 8.43 | 7.97 | 8.19 |

**Table SI7.** The actual and the calculated multiplicities for two genome pairs in the case of random fluctuations. N - genome number; AM - actual multiplicity, 10, 20,...,50 - segment lengths. In the case of the first pair, the actual multiplicity cannot be calculated (-3.01 as compared to 4 and 8.74 as compared to 3). However, the sums of the actual (7) and calculated (5.73) multiplicities are much closer. For all the mixtures, the sum of the obtained multiplicities equals approximately 6. Similarly, for the second pair, the difference between the actual and the calculated multiplicities is much larger than the difference between the corresponding sums (9 for the actual and about 9 for the calculated multiplicities).

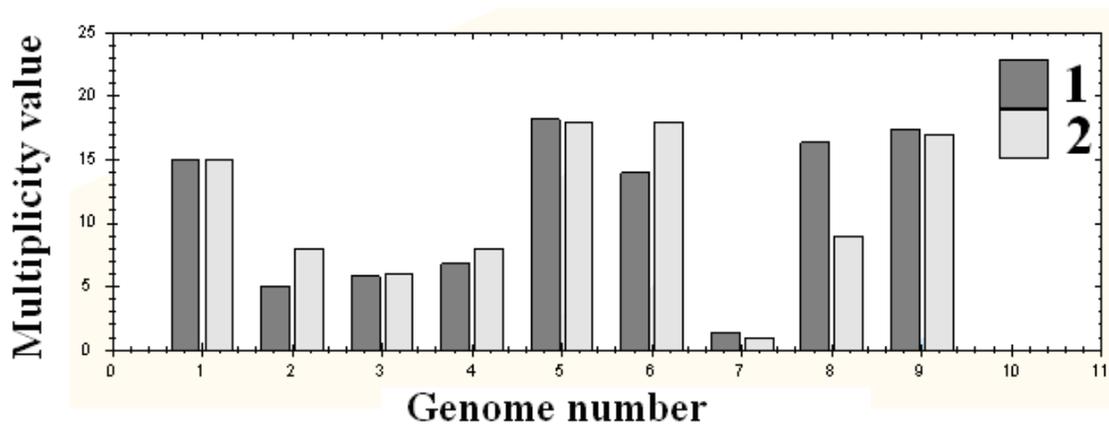

**Fig. SI1.** Mixture of nine genomes: **1** - *Campylobacter1 jejuni*; **2** - *Salmonella*; **3** - *Pseudomonas aeruginosa*; **4** - *Vibrio cholerae*; **5** - *Mycobacterium tuberculosis*; **6** - *Escherichia coli*; **7** - *Legionella pneumophila*; **8** - *Shigella boydii*; **9** - *Yersinia enterocolitica*. **(1)** - actual multiplicities; **(2)** – multiplicities calculated based on the 10-letter vocabulary (200 words with 3 mismatches.